\documentclass[10pt, sigconf, nonacm]{acmart}
\usepackage{graphicx}
\usepackage{xcolor}
\usepackage{url} 
\usepackage{appendix}
\usepackage{todonotes}
\newcounter{todocounter}
\usepackage[caption=false]{subfig}
\usepackage{amsmath}
\usepackage[caption=false]{subfig}
\usepackage{tikz}
\definecolor{myorange}{RGB}{255,140,0}
\hypersetup{linkcolor=[rgb]{.75,0,0},urlcolor=[rgb]{0.75,0,0.75},citecolor=blue}
\def\Snospace~{\S{}}
\newcommand{\RelaxFloats}{
	\renewcommand{\topfraction}{0.9}
	\renewcommand{\floatpagefraction}{0.9}
	\renewcommand{\textfraction}{0.1}
}
\RelaxFloats

\makeatletter
  \newcommand\EatSpacesHack{\@bsphack\@esphack}
\makeatother

\makeatletter
\let\comment\@undefined
\let\endcomment\@undefined
\makeatother

  \newcommand\comment[1]{\textbf{\sffamily\color{blue}[xxx: #1]}}
  \newcommand\oldreviewfix[1]{\EatSpacesHack}

  \newcommand\PostSubmission[1]{\EatSpacesHack}
  
  \makeatletter
    \renewcommand\section{\def\@toclevel{1}%
      \@startsection{section}{1}{\z@}%
      {-.3\baselineskip \@plus -2\p@ \@minus -.2\p@}%
      {.15\baselineskip}%
      {\ACM@NRadjust\@secfont}}
    \let\ACM@origsection\section
    \renewcommand\subsection{\def\@toclevel{2}%
      \@startsection{subsection}{2}{\z@}%
      {-.2\baselineskip \@plus -2\p@ \@minus -.2\p@}%
      {.15\baselineskip}%
      {\ACM@NRadjust\@subsecfont}}
    \let\ACM@origsubsection\subsection
    \renewcommand\subsubsection{\def\@toclevel{3}%
      \@startsection{subsubsection}{3}{\z@}%
      {-.2\baselineskip \@plus -2\p@ \@minus -.2\p@}%
      {-3.5\p@}%
      {\ACM@NRadjust{\@subsubsecfont\@adddotafter}}}
    \renewcommand\paragraph{\def\@toclevel{4}%
      \@startsection{paragraph}{4}{\parindent}%
      {-.5\baselineskip \@plus -2\p@ \@minus -.2\p@}%
      {-3.5\p@}%
      {\ACM@NRadjust{\@parfont\@adddotafter}}}
    \let\ACM@origsubsubsection\subsubsection
    \renewcommand\noindentparagraph{\def\@toclevel{4}%
      \@startsection{paragraph}{4}{\z@}%
      {-.5\baselineskip \@plus -2\p@ \@minus -.2\p@}%
      {-3.5\p@}%
      {\ACM@NRadjust{\@parfont}}}
  \makeatother

\begin{document}
\raggedbottom

\title{Bounding Mobile Latency from Network Interconnections}

\author{Shefali Tushar Kulkarni}
\affiliation{
  \institution{USC/Information Sciences Institute}
  \city{Los Angeles}
  \country{United States}}
\email{shefalit@usc.edu}

\author{John Heidemann}
\affiliation{
  \institution{USC/Information Sciences Institute}
  \city{Los Angeles}
  \country{United States}}
\email{johnh@isi.edu}

\begin{abstract}
Mobile latency is determined by how Mobile Network Operators (MNOs) route traffic: 
from phones, across the MNO's network
to \emph{exit points} where it enters the public Internet.
Exit points are few, and MNOs are quiet about details,
  but their role prompts us to ask:
What is the latency cost of send traffic through an exit point?
We develop a new method to evaluate this question,
  and demonstrate it on the three largest U.S. MNOs,
  although in principle it can apply globally.
Using new analysis of existing traceroutes
  we find that AT\&T, 
  T-Mobile, and Verizon funnel nationwide traffic through only 9, 5, and 9 exits,
  and the geographic detour they incur adds
  up to 62.1\,ms compared to an ideal, direct-path latency. 
Routing delays also add latency,
  we show that 
  even assuming speed-of-light paths through the exit points,
  latency is still up to 46.9\,ms more than optimal.
To understand if these costs can be overcome
  we compare actual to theoretically optimal placements,
  finding that latency of 
  current placements are 18--66\% above the population-weighted optimum
  and 32--72\% above the market-share-weighted with the same number of exits.
However, we show this gap is small in absolute terms:
  perfect placement would recover only a few milliseconds
  of the tens that remain.
We show that geographic optimization saves little latency:
  the inflation users experience comes from routing and other overhead.
\end{abstract}



\maketitle

\section{Introduction}

\begin{figure}
    \centering
    \mbox{\includegraphics[width=1.0\columnwidth,trim=0 150 50 140,clip]{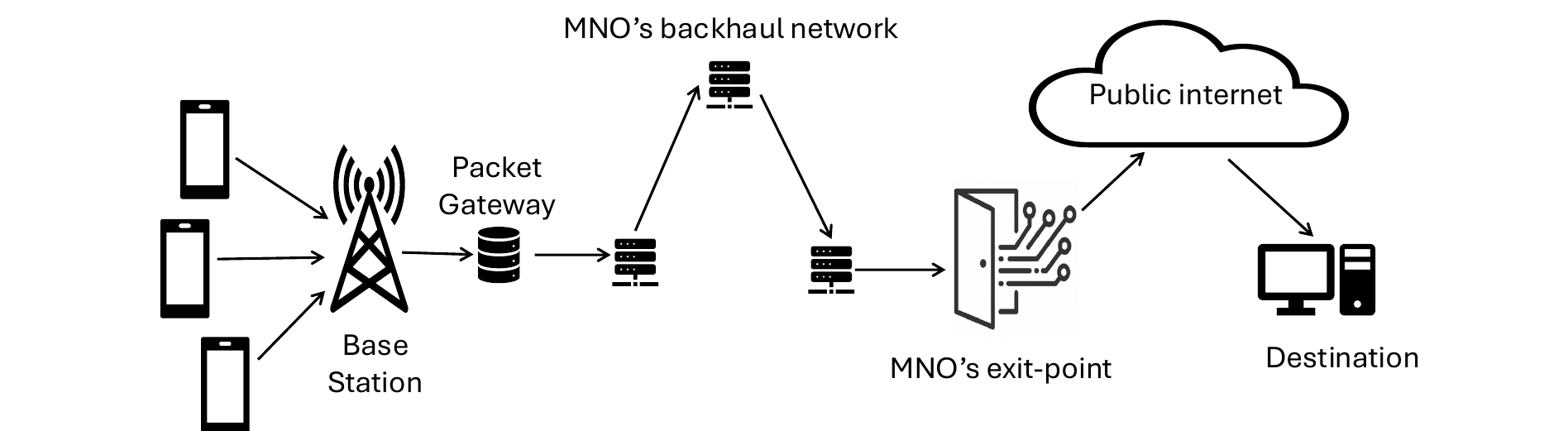}}
    \caption{The path from a mobile device, through the PLMN backhaul and exit point to the Internet.}
    \label{fig:PLMN_network}
\end{figure}

Today, 
phones dominate human access to the Internet. 
Individuals are continuously connected, 
  and expect fast network access wherever they are.
A mobile phone connects to the Internet via the Public Land Mobile Network (PLMN)
  of the Mobile Network Operator (MNO). 
The mobile phone uses a 4G or 5G Radio Access Network (RAN), Packet Gateways, Edge,
  and traffic goes through the PLMN's backhaul
  to an \emph{exit point} that interconnects to the public Internet~(\autoref{fig:PLMN_network}).

Large online services and CDNs co-locate servers within PLMN data centers and at IXPs near exit points~\cite{Datacenters,GoogleDatacenters,MetaDatacenters}
to reduce the distance traffic travels after leaving the mobile network.
Mobile traffic must still traverse the PLMN's backhaul to reach these servers,
and a large-scale study finds that bringing data centers closer
to its users does not guarantee low end-to-end latency~\cite{Zhang19b}.
Our paper is the first to explore the question of
  \emph{how much latency is added by detours created by the MNO backhaul network and exit points?}
We answer this question by locating the current exit points of three U.S. carriers: AT\&T, T-Mobile, and Verizon.

We build on two prior measurements of U.S.~carriers.
Researchers at UCSD mapped the internal topology of the same three largest U.S. MNOs,
  focusing on \emph{interior} the PLMN's network in 2021.
They highlight the role of Packet Gateways, %
  that bridge the dedicated wireless backhaul to
  the MNO's internal backbone~\cite{Zhang21b}.
We instead focus specifically on the \emph{exit point
  at the boundary between the MNO and the public Internet}.
We use their dataset but go beyond their analysis
  to explore the number and placement of exit points.
Researchers at Northeastern independently collected traceroutes
  studying the same U.S.~carriers in 2024~\cite{ghoshal2025replication}.
They emphasize 5G application and radio performance.
We apply our new analysis to their data to validate exit points in 2024.

Our first contribution is we develop a new method
  that isolates delay
 required by physical distances
  from delays added by routing.
We define three paths from a handset to a destination:
  the \emph{observed path}, measured end-to-end;
  the \emph{direct path} at speed-of-fiber directly between source and destination;
  and the \emph{exit path} at speed-of-fiber through the operator's
  own nearest exit point (\autoref{sec:addtional_latency_methodology}).
The difference between observed and direct path gives
  \emph{additional direct-path latency};
  the difference between observed and exit-path gives
  \emph{additional exit-path latency}.
Computing additional latency requires knowing where the current exit points are,
  so our first step is a new method to identify exit points from traceroutes
  (\autoref{sec:exit_point_methodolgy}).
We  find 9 exit points for AT\&T, 5 for T-Mobile, and 9 for Verizon
  (\autoref{sec:result_exit_points}).

Second, we apply this method to the three largest U.S. MNOs.
(In principle our method can apply to non-U.S. MNOs as well.)
Paths through these exits give a median of 62.1\,ms (AT\&T),
  43.9\,ms (T-Mobile), and 49.1\,ms (Verizon) additional direct-path latency (\autoref{sec:result_geo_latency}).
We attribute this residual to hot-potato routing:
  operators hand traffic off at their nearest exit
  rather than the one nearest the destination,
  a known cause of path inflation~\cite{Spring23}
  that has not been quantified for U.S. mobile networks.
Granting each MNO the exits it already operates,
  a median of 23.6\,ms, 37.2\,ms, and 46.9\,ms remains
  that distance cannot explain (\autoref{sec:result_exit_latency}).
Most of the delay is therefore not the geographic detour through exit points but, 
we argue, 
routing policy.

Our final contribution examines how well MNOs have placed their exit points:
  do they account for proximity to users?
  could they do better?
We formulate placement as a facility-location problem
  over population density and over population weighted by MNO market share
  (\autoref{sec:optimizing_location_methodology}).
AT\&T, T-Mobile, and Verizon exceed the optimum by 18\%, 66\%, and 43\%,
  and by 32\%, 61\%, and 72\% under market-share weighting.
AT\&T and Verizon place most exits near optimal,
  with the gap coming from a few outliers;
  T-Mobile's larger gap reflects its use of third party backbone.
Since our optimum ignores cost, peering, and available physical paths,
  it bounds what any operator could achieve.
We show that after a service has five well-placed exits,
  additional exits have minimal benefits, reducing
  mean latency, per capita, by under 1\%
  (\autoref{sec:exit_optimality}).
  
The impact of our work is to provide operators and regulators a way to evaluate exit-point placement
  and reason about the trade-offs of adding new exits.
Our approach extends to other regions and operators.

Our work analyzes traceroute data from 2021~\cite{Zhang21b},
  with validation against an independent collection from 2024~\cite{ghoshal2025replication},
  so it poses no ethical concerns about user privacy. 
While this data is now 5 and 2 years old,
  we show our results are consistent over time (\autoref{sec:validation}).
Our basic data is available to researchers~\cite{Zhang21b,Ghoshal25ReplicationData,Ghoshal25b},
  and we will make the data from our analysis available at no cost~\cite{Kulkarni26c}.

This paper is published at the 21st Asian Internet Engineering Conference (AINTEC)
  in 2026 in Taipei, Taiwan~\cite{Kulkarni26a}.
This technical report adds appendices
  with additional information about distance and latency (\autoref{app:distance_latency})
  and evaluating the latency improvement from additional exits (\autoref{app:optimizing_n+i_location}).

\section{Related Work}

We build on prior studies of PLMNs,  mobile latency,  geolocation,  topology  optimization,  and facility location.

\emph{Mobile network studies:}
Most closely related to our work are previous studies on the wide-area performance of PLMNs.
Zhang \emph{et al.}~collected traceroutes from mobile phones physically mailed around the U.S.,
  creating the \emph{ShipTraceroute} dataset,
  and used it to study the router-level topology of the backhaul networks
  of three U.S.~MNOs~\cite{Zhang21b}.
   Their analysis focuses on the internal network
 of the PLMN, from the base station, through the packet gateway (PGW) and the operator's backbone, to the exit point.
 Our analysis picks up where they leave off, looking at how that exit point
   affects end-to-end latency (\autoref{sec:latency_model}),
   and further exploring the impact of number of exit points.
For latency analysis, they report only minimum RTT to the destination;
  we add theoretical optimal latency to show latency inflation.
  
A second study uses data from human driven mobile phones
  to evaluate coverage, geodiversity, operator diversity, throughput, RTT, and application performance~\cite{Ghoshal23}.
  We use the traceroutes from their later replication of this drive
  to validate exit-point locations observed in 2024~\cite{ghoshal2025replication}.

Many studies characterize commercial mobile performance in the wild,
  measuring throughput, latency, and device behavior across U.S.~cities~\cite{ghoshal2022depth,narayanan2020first,narayanan2020lumos5g,narayanan2022comparative,narayanan2021variegated},
  in single cities outside the U.S.~\cite{kousias2022implications,kousias2022coverage,xu2020understanding,Zmyslowski24},
  at scale through an app~\cite{yang2022mobile},
  from the operator's perspective~\cite{parastar2023spotlight},
  and under mobility~\cite{deng2018mobility,ni2023polycorn,wang2019active}.
  Zarifis \emph{et al.}~diagnose mobile path inflation
  and conclude it had disappeared in the U.S.~by 2012~\cite{zarifis2014diagnosing};
  we find it persists a decade later.
At global scale, Varvello \emph{et al.}~find that only 20\% of 24 mobile networks achieve sub-20\,ms latency~\cite{varvello23},
  Ozcan \emph{et al.}~show from 9.6 billion RIPE Atlas pings that inter-city delays have generally improved,
   in North America and Europe~\cite{ozcan23},
  and a year-long study of eight cities finds that 5G has stagnated
  in delivering consistent latency improvements over LTE~\cite{khan25}.
Together these studies provide a rich view of cellular performance and mobility,
  but they do not examine how core-network topology and exit-point placement constrain wide-area mobile latency.
We instead infer the number and placement of MNO exit points,
 examining if limited exit points keeps mobile users from benefiting from fiber and RAN-level improvements.

Latency matters most for interactive applications:
  K\"am\"ar\"ainen \emph{et al.}~measure latency in mobile video and gaming,
  reporting 20\,ms of network delay under optimal conditions
  and showing the effect of geographic distance~\cite{Kaemaeraeinen17},
  while work on Mobile/Edge Cloud places computation near the handset
  to cut latency for cloud games~\cite{Zhang19b}.
Such applications increasingly rely on CDNs to reduce latency~\cite{frangoudis2017cdn},
  but CDN nodes usually sit just outside the MNO's network,
  and some requests must still reach regional cloud datacenters---in both cases,
  exit-point count and placement affect latency.
Recent studies infer physical network boundaries
  (submarine cables) from active measurements~\cite{Alagappan23a, wangcalypso}.
We apply similar measurements to a different boundary:
  private-to-public exit points within the U.S.~mobile network, whose locations, unlike cable landing stations, are not public.

\emph{Optimal latency and path inflation in wide-area networks:}
Singla \emph{et al.}~compare observed Internet latency against the physical limits of fiber propagation,
  framing a ``speed-of-light'' Internet as the ultimate goal~\cite{Singla14b}.
Latency is also the defining advantage in high-speed trading,
  motivating the use of microwave links instead of fiber to reduce delay~\cite{Shkilko20a}.
We also measure PLMN latency against speed-of-fiber limits.
The exit-point-induced latency we study is a specific instance of \emph{path inflation},
  where non-direct routing increases delay.
Path inflation has been widely studied in wireless networks,
  showing that poor routing can significantly inflate latency in IP anycast services~\cite{Li18a, Spring23},
  and that while such inflation may be acceptable for some applications,
  including the DNS root even at delays above 100\,ms,
  it remains critical for latency-sensitive services such as CDNs~\cite{Koch21}.
Consistent with these findings,
  we observe comparable latency inflation in our exit-point analysis (\autoref{sec:result_exit_latency}).
Sundberg \emph{et al.}~analyze over billions of RTT samples from a wireless ISP using eBPF,
  showing internal network latency exhibits large, time-varying tails during busy hours~\cite{sundberg24}.
While their work focuses on last-mile variability,
  we test the middle-mile network.

\emph{Exit-point placement:}
We formulate optimal exit-point placement as a facility location problem~\cite{FacilityProblem,uncapicitated83,liu09},
  in which facilities (such as warehouses) are placed to minimize travel time.
K-means and K-Medians have long been recognized as good, approximate solutions to this problem~\cite{Bradley00a,jesus2012figures,Feng20},
  and we build on this method to evaluate optimal exit placement (\autoref{sec:optimizing_location_methodology}).
Recent work shows that large-scale placement problems can be solved efficiently
  with provable performance bounds using LP-relaxation heuristics~\cite{behravesh24},
  and that optimal placement in wireless mesh networks is NP-hard~\cite{Bing07}.
Unlike wireless-mesh networks, which are often decentralized and ad hoc,
  mobile network infrastructure is fixed and carefully planned to provide good coverage,
  so we use population-driven and market share measurements to quantify how far real-world MNO exit deployments
  deviate from the theoretical optimum (\autoref{sec:exit_optimality}).

\section{Methodology}
    \label{sec:methodology}

We next develop a new method that
  identifies two parts of the mobile latency:
  latency due to geography, and due to routing.
We then apply it to AT\&T, T-Mobile, and Verizon.

We define this problem (\autoref{sec:problem_statement}),
  types of latency (\autoref{sec:latency_model}),
  and data sources (\autoref{sec:datasets}).
We then describe our new methodology for analyzing traceroute data
  to
  identify exit points and their locations (\autoref{sec:exit_point_methodolgy}),
  quantify additional latency (\autoref{sec:addtional_latency_methodology}),
  and evaluate optimality with fixed and additional exit points
  (\autoref{sec:optimizing_location_methodology}).

\begin{figure}
    \centering
    \includegraphics[width=0.8\columnwidth]{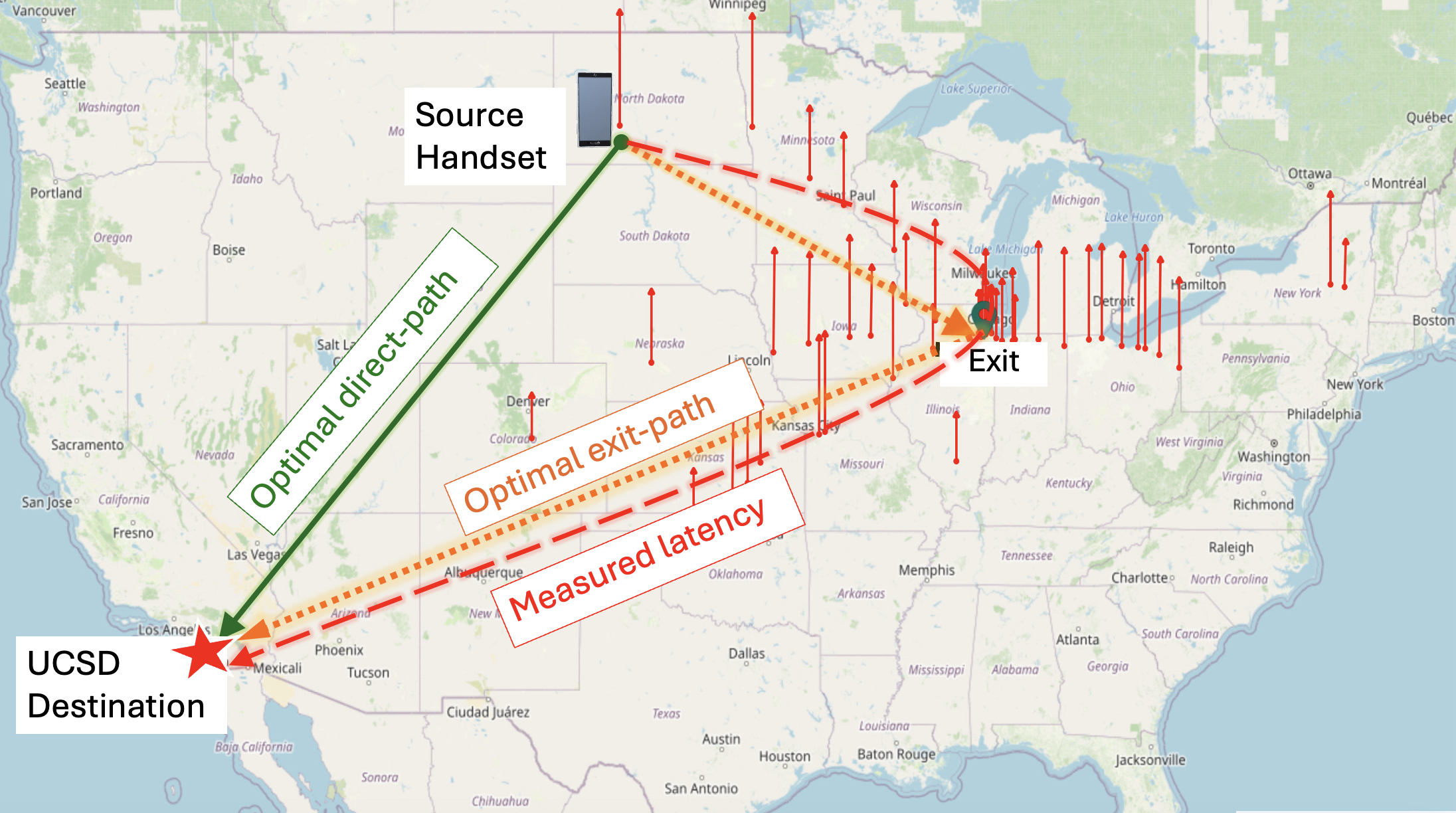}
    \caption{Additional latency from the handset to the Internet: observed (red dashed), optimal direct-path (green solid), optimal exit-path (orange dotted). Vertical red lines give additional direct-path latency through AT\&T's Chicago exit.}
    \label{fig:latency_methodology}
\end{figure}

\subsection{Problem Statement}
	\label{sec:problem_statement}

In a Public Land Mobile Network (PLMN), 
traffic from a mobile device reaches the public Internet only after traversing the operator’s backhaul network and exiting through a limited number of exit points, as shown in \autoref{fig:PLMN_network}.
In that backhaul network
  the PLMN controls the path and exit point. 
After traffic leaves the their AS, 
external networks determine the subsequent path according to their own BGP policies.
Because exit-point locations are not public,
  several important questions remain open:
How much additional latency is caused because of geographic detour?
How good is exit-point placement, and what placement would optimize latency?
How much will adding new exit points improve latency?
Being able to reason about these questions can help us evaluate
  how well a PLMN network architecture
  serves its customers.

\subsection{Latency and Definitions}
\label{sec:latency_model}
To understand how PLMN architecture affects user latency
  we define several components of that latency.
  
We define three paths between the user and a destination.
The \emph{direct path} is the great-circle path from the mobile device to the server,
  traversed at the speed of light in fiber, roughly two-thirds of its speed in vacuum.
In \autoref{fig:latency_methodology}, the direct path is the green line
  from North Dakota to San Diego.
The \emph{exit path} follows the same assumptions,
  but routes from mobile device through the PLMN's nearest exit point to the server---it is the orange path from North Dakota via Chicago to San Diego.
The \emph{observed path} is the path taken by the traceroutes,
  the slightly longer dotted red line.

These paths identify overhead types.
\emph{Additional direct-path latency} is the difference
  between observed and optimal direct-path latency.
The optimal direct-path is the minimal physical distance from source to destination,
  so this difference observes latency to due detouring through an exit-point.
Second, we define  
  \emph{additional exit-path latency} as the difference
  between observed and optimal exit-path latency.
Here optimal exit-path latency is the physical distance
  from source to exit-point to destination,
  so this latency isolates latency from factors \emph{other} than this detour.
The optimal paths always assume physical distances only,
  so they ignore per-router network processing delays
  and extra latency when physical paths are not ``as-the-crow-flies'',
  so these latencies are optimistic.
These bounds are useful baselines
  because the gap between observed and bound is the overhead we set out to measure~\cite{Singla14b}.

Finally, we ask how well the current exit points are placed
  by solving for the optimal placement of a given number of exits.
Real deployments are constrained:
  fiber follows a limited set of physical paths~\cite{Durairajan15a},
  and data centers have fixed locations.
Those constraints can only move placement away from the optimum,
  so our solution provides an upper-bound
  on
  the latency any operator could recover by relocating exits.

\subsection{Datasets}
    \label{sec:datasets}

Our work draws on several key datasets:
  ShipTraceroute and the Northeastern dataset for paths from mobile phones to the Internet,
  OpenCellID to geolocate mobile phones,
  IP addresses allocations from ARIN to identify network ownership,
  IP2Location to geolocate IPs,
and finally census data for the U.S.~population by county.
We describe each below.

\subsubsection{Mobile Network Traceroutes}

ShipTraceroute~\cite{Zhang21b} is our primary source for known PLMN
topology and latency. 
It consists of hourly traceroutes from three
phones, 
one per U.S. MNO (AT\&T, T-Mobile, and Verizon),
mailed to
locations across 40 states in the continental U.S. in 2021. 
Targets include
servers at UCSD, Google,
Netflix, and other peers. 
We identify
exit points from the Google, 
Netflix, and UCSD traceroutes, and
measure additional latency from the UCSD data alone
(\autoref{sec:exit_point_methodolgy}). 
Because GPS is often unavailable, we locate phones by ShipTraceroute's recorded cell tower IDs.

To validate exit points in 2024 (\autoref{sec:validation}), 
we use traceroutes from a second, independent mobile dataset~\cite{ghoshal2025replication}. 
It covers same three MNOs with fewer samples.

\subsubsection{Mobile Locations}
Knowing where each traceroute originates is essential to measuring
source-to-destination latency. Since GPS is often unavailable, 
we map
the cell tower IDs in ShipTraceroute to locations using
OpenCellID~\cite{OpenCellID}, a database of tower locations, Cell IDs,
and PLMN metadata. 
IP geolocation was considered but discarded, 
as MNOs typically use private IP space, preventing accurate geolocation.
\subsubsection{IP to AS Mapping}

We detect network transitions along traceroutes using the ISI outage
supplement dataset~\cite{ant_block_to_as},
a pre-built map from each
/24 prefix to its ASN, aggregated from the five global RIRs. 
A
transition out of the MNO's AS marks an exit point. 
We use /24
granularity as it is the finest prefix commonly routed and observed in
global BGP.

\subsubsection{IP2Location}

We determine the physical locations of PLMN exit points by combining
IP2Location geolocation with IP-to-AS mappings. Geolocation accuracy
remains an active research area. We considered MaxMind GeoLite,
GeoIP2, and IP2Location-Lite~\cite{IP2LOCATION}; prior work on Internet
infrastructure finds their accuracy comparable, particularly in the
United States~\cite{Livadariu20, Gharaibeh17, Yasin2023}.
  We chose IP2Location-Lite for its slightly better city-level coverage.
(We reconfirmed this with a mini-study of 200 known, U.S. university locations.)
Although city‐level geolocation accuracy can err by tens of kilometers, 
our analysis concerns latency shifts on the order of tens of milliseconds. 
Since light in fiber travels roughly 200 km per millisecond, 
city‐level accuracy is more than adequate for detecting the effects we observe.

\subsubsection{Population Estimation}
We use the U.S.~Census data from year 2023 to locate the U.S.~population by county~\cite{PopulationDensity} 
and state-level MNO market share~\cite{marketshare}
  as a proxy for each operator's subscribers.
We use this data to evaluate the optimality of exit-point placement
  in \autoref{sec:optimizing_location_methodology}.

\subsection{Identifying the exit points}
	\label{sec:exit_point_methodolgy}

Our first objective is to identify exit points: the locations where
traffic leaves the PLMN for the public Internet.

To identify these exit points, 
we extract all traceroutes for an MNO from the ShipTraceroute dataset and determine the network operator of each valid hop using address-to-AS data. 
An exit point is identified as the first hop where the traceroute transitions from an MNO's IP to a public IP.
In most network architectures, 
the first public IP encountered marks the interface where routing responsibility shifts from the MNO to an external operator. 
By verifying that the Autonomous System associated with this IP is not part of the MNO, 
we distinguish between internal and external routing.

We analyze 1,592 successful traceroutes for AT\&T, 
1,720 for Verizon, 
and 872 for T-Mobile from the ShipTraceroute dataset. 
Although the total number of traceroutes is limited, 
our analysis aggregates measurements at the level of base stations and exit points, 
and we observe stable exit-point inference and latency estimates with relatively few samples per site. 
Per MNO, 
we find 9 exit points for AT\&T and Verizon. 
Prior work reports that T-Mobile uses multiple backbone providers that serve each region at different interconnection points~\cite{Zhang21b}. 
Consistent with this, 
we observe that T-Mobile's traceroutes to the UC San Diego server traverse Zayo (AS 6461) and Lumen (AS 3356), 
which provide Internet transit service to T-Mobile. 
We therefore provide a second look at the T-Mobile–backbone combination and identify 5 exit points under this model (details in \autoref{sec:result_exit_points}).

\emph{Role of the Destination:}
Exit points depend on the source and destination locations.
We consider traceroutes with a single source: 
the handset.
ShipTraceroute has traceroutes to three destinations: UCSD, Netflix, and Google.

For AT\&T and Verizon we infer exit-point locations from traceroutes to all three destinations.
For T-Mobile, we use only traceroutes to UCSD,
  because we find traceroutes to Google and Netflix from T-Mobile (AS 21928) terminate within T-Mobile's network in different internal locations
    due to a practice known as \emph{Off-Nets}~\cite{Gigis21a}.

Although exit-point determination can use all three destinations,
  in the next section we discuss that estimation of additional latency
 requires a single destination.

\subsection{Quantifying Additional Latency}
	\label{sec:addtional_latency_methodology}

Given the identified exit points, 
our next step is to assess their impact on latency. 
As described in \autoref{sec:latency_model}, 
our goal is to compare \emph{measured} latency with both the optimal direct-path latency and the optimal latency via the nearest exit point.
We determine the known latency from traceroutes. 
Traceroute records the round-trip time (RTT)
  taken to each hop from the cellphone to a destination,
  with the last hop showing the end-to-end latency~\cite{Traceroutes}.

We approximate the cellphone location for the origin of the traceroute.
While GPS data is unavailable, 
ShipTraceroute dataset provides the Local Area Code (LAC) of the serving cell tower at the time of the traceroute. 
We map LAC to physical location using OpenCellID. 
Using OpenCellID metadata, we map each traceroute to its corresponding base station.
We use the serving base station's location
  as a proxy for the handset's location when calculating theoretical latency; 
the RTT itself is measured end-to-end from the handset.

Our goal is to estimate propagation latency
  (roughly speed-of-fiber for physical distance)
  while ignoring queueing delay and other transient effects.
We estimate propagation latency 
  as the minimum
  RTT of all observations from the same device and location,
  following TCP Vegas' observation that minimum latency
  reflects physical propagation delay~\cite{Brakmo95a}.

To keep analysis of additional latency consistent,
  we consider only traceroutes to one known destination (San Diego),
  and from known (geolocatable) source locations.
Although ShipTraceroute also measures paths to Netflix and Google, 
we exclude these destinations
  because computing optimal latency uses the destination’s
  physical location. 
With anycast and load balancing,
  the location of Google and Netflix is unknown.
   The exit points we find are the same for any destination, and only the size of the residual changes.
The choice of measurement to a single location on the U.S.~west coast 
  will affect the relative size of detours--small detours
  may be a larger relative change for users on the west coast,
  but are washed out by the higher absolute propagation latency for east-coast users.
Although bias is a limitation of ShipTraceroute
  and implies our absolute results should be considered preliminary,
  our methodology of comparing actual and optimal latency generalizes.

For optimal latency, we estimate the round-trip time (to and from the destination) over the shortest physical fiber path, assuming propagation at the speed of light in fiber.
The speed-of-fiber, approximately 200~km/ms (about two-thirds the speed of light), 
determines data propagation through fiber optic media. 
We calculate the great-circle distance
  as shortest possible physical distance
 from cellphone to the fixed destination. %
As a theoretical optimal, we ignore limitations of physical fiber paths.

To compute the best possible exit-path latency, 
we identify the exit point from the traceroute (where traffic leaves the PLMN, 
as per \autoref{sec:exit_point_methodolgy}) and determine its physical location using IP2Location. 
We then compute the exit-constrained physical distance as the sum of the distances from the handset to the exit point and from the exit point to the destination in San Diego. 
The estimated latency is obtained by applying the speed-of-fiber to this total distance.

Using the measured latency along with optimal direct and exit-path latency, 
we compute: \emph{additional direct-path latency} as the difference between known latency and optimal direct-path latency, 
and \emph{additional exit-path latency} as the difference between known latency and optimal exit-path latency.

\subsection{Determining Optimal Exit Locations}
	\label{sec:optimizing_location_methodology}

Finally, given what we learn about each MNO's network,
  we compare their known exit-point locations to the 
  optimal possible exit-point locations.
This estimate complements our known latency measurements
  and their overhead (\autoref{sec:addtional_latency_methodology})
  to find the best-possible latency for the continental U.S.~given
  a limited number of exit points.

Given $n$ exit points, we define their cost as
  the distance from every person in the continental U.S.,
  through their nearest exit point, to San Diego.
The optimal location for a fixed number of $n$ 
  exit points minimizes this objective.
Optimizing the location of $n$ sites is the \emph{facility location problem}~\cite{FacilityProblem, eiselt2011pioneering, marianov2011median}.
While an optimal solution is \emph{NP}-complete,
  $k$-means provides a well-known approximate solution~\cite{Bradley00a}.

We therefore estimate optimal exit-point location for each MNO
  by taking their known number of exit points as $k$,
  using county-by-county U.S.~population data from the Census Bureau
  to optimize cost using $k$-means.
We optimize exit-point placement using population as a proxy for expected demand,
  assuming more populous counties contribute more traffic.
  More granular metrics like per-cell traffic volume or MNO-specific subscriber distributions would be ideal, 
  but are not publicly available.
  Population data provides a consistent, 
  accessible basis for optimization across MNOs.

Our prior analysis fixes $n$ at each MNO's known count,
  but in practical systems there is often a temptation
  to over-provision infrastructure
  on the assumption that \emph{more is better}.
To test that assumption for exit points,
  we add exits incrementally beyond the optimal set,
  placing each by $k$-means over the remaining population.
This measures the marginal benefit of exit-point density
  under idealized conditions (\autoref{sec:exit_optimality}).

\subsection{Limitations}
    \label{sec:Limitations}
Our analysis is subject to limitations in
  destination diversity,
  measurement times,
  and routing visibility.
Although we measure latency only to UCSD, our method generalizes to other destinations, as discussed in detail in \autoref{sec:addtional_latency_methodology}. 
Our analysis relies on ShipTraceroute measurements reported in 2021. 
Comparison with Northeastern data from 2024 provides evidence of exit-point stability at at least few of the locations examined,
supporting the continued relevance of our analysis. 
Further details are discussed in \autoref{sec:validation}.
Our traceroute-based analysis cannot directly verify routing policies, 
distinguish the effects of policies within the MNO from those beyond the exit point,
or identify interconnection types.
We also considered RouteViews BGP data to assess exit-point and route stability over time, 
but RouteViews does not host the relevant mobile routes.
Future work could address these limitations through operator-provided routing and interconnection data.

\section{Results}
	\label{sec:results}

We next %
  identify exit points (\autoref{sec:result_exit_points}),
  how exit location affects latency (\autoref{sec:result_geo_latency} and \autoref{sec:result_exit_latency})
  and optimality (\autoref{sec:exit_optimality})
  in 3 MNOs.
  
\subsection{Exit-Point Locations}
	\label{sec:result_exit_points}

We first report the number of observed exit points for each mobile network.
\autoref{fig:exit_point_locations} presents the number and locations of exit points for each MNO. 
We find that T-Mobile operates 5 exit points, 
while AT\&T and Verizon each have 9 as of 2021. 
These exit points are distributed across the continental U.S., a topic further examined in \autoref{sec:exit_optimality}.
Are these exit points sufficient? 
On one hand, these MNOs remain profitable, 
and their customers appear satisfied, suggesting that they are adequate.
However, 
Internet-centric companies prioritize broad interconnection; Google, for instance, 
peers at 25 sites in the continental U.S. as of July 2024~\cite{PeeringDBGoogle24a} and deploys edge nodes within thousands of ISPs~\cite{Calder13a}. 
While the number of exit points is ultimately a business decision for each MNO, 
it remains critical to understand how those choices affect users. In our next sections, we measure the end-to-end delay caused by the sparse exit-point topology and analyze where additional exit points can deliver most  latency reduction.

\subsection{Additional Direct-Path Latency}
	\label{sec:result_geo_latency}

Having identified PLMN exit points, 
we now examine \emph{additional direct-path latency}, 
defined in \autoref{sec:latency_model} as the difference between observed latency and the theoretical direct physical path from the mobile phone to its Internet destination.

\begin{figure*}[t]
\centering
\subfloat[AT\&T: 9 exits: Seattle, San Francisco, Los Angeles, Dallas, Chicago, Atlanta, Ashburn, New York, Boston; and add. direct path latency.]{
  \begin{tikzpicture}
    \node[anchor=south west, inner sep=0] (img) at (0,0) {
      \includegraphics[width=.32\textwidth]{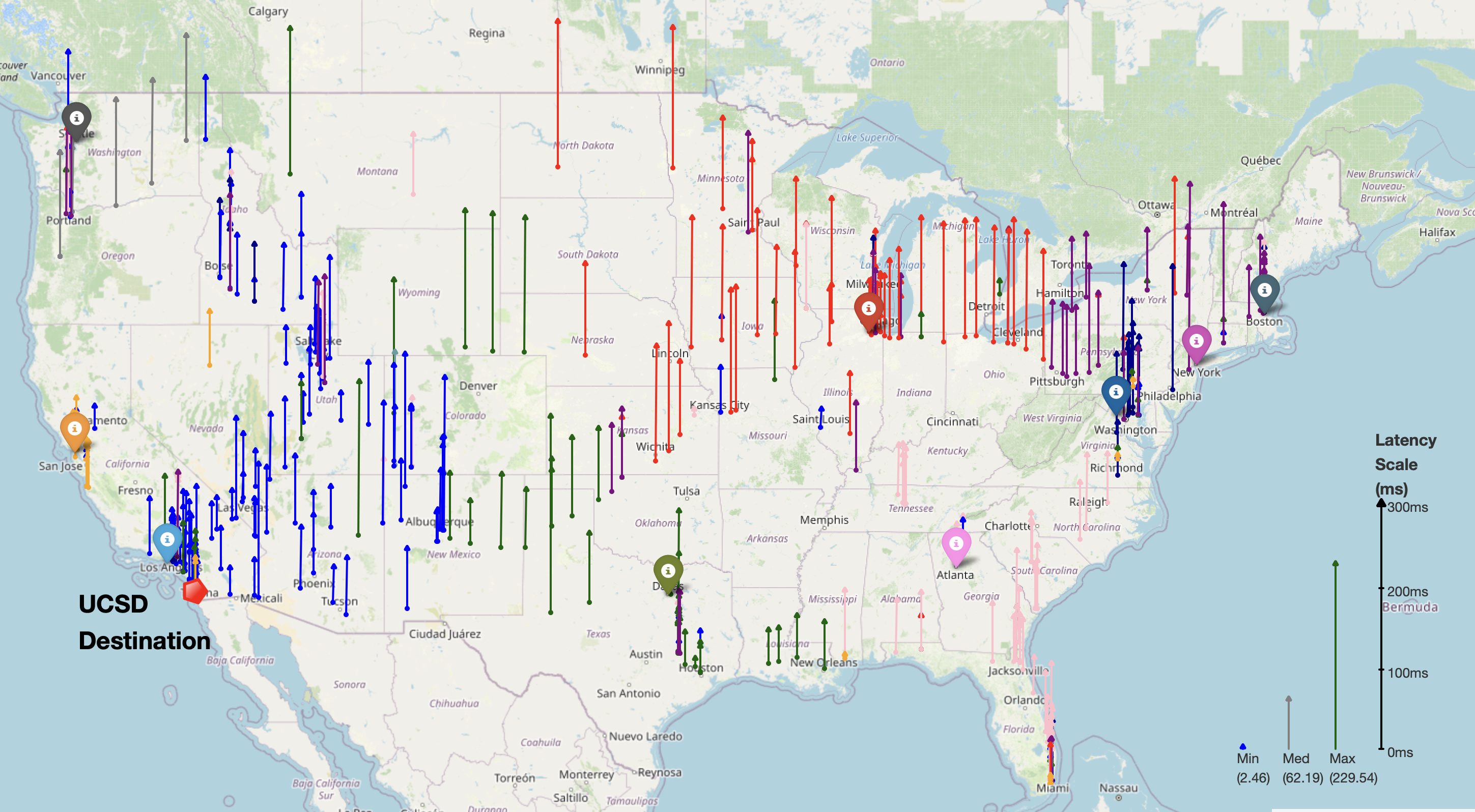}
    };
    \begin{scope}[x={(img.south east)}, y={(img.north west)}]

      \draw[thin, dashed] (0.37, 0.64) rectangle (0.49, 0.88);
      \node[font=\scriptsize\bfseries, align=center] at (0.44, 0.795) {Dakotas};

    \draw[thin, dashed] (0.22,0.33) rectangle (0.36,0.50);
\node[anchor=center, font=\scriptsize\bfseries, align=left] 
  at (0.27,0.405) {Four\\Corners};

    \end{scope}
  \end{tikzpicture}
}\hfill
\subfloat[T-Mobile: 5 exits: Cupertino, Los Angeles, Phoenix, Denver, Chicago; and add. direct path latency.]{
  \begin{tikzpicture}
    \node[anchor=south west, inner sep=0] (img) at (0,0) {
      \includegraphics[width=.325\textwidth]{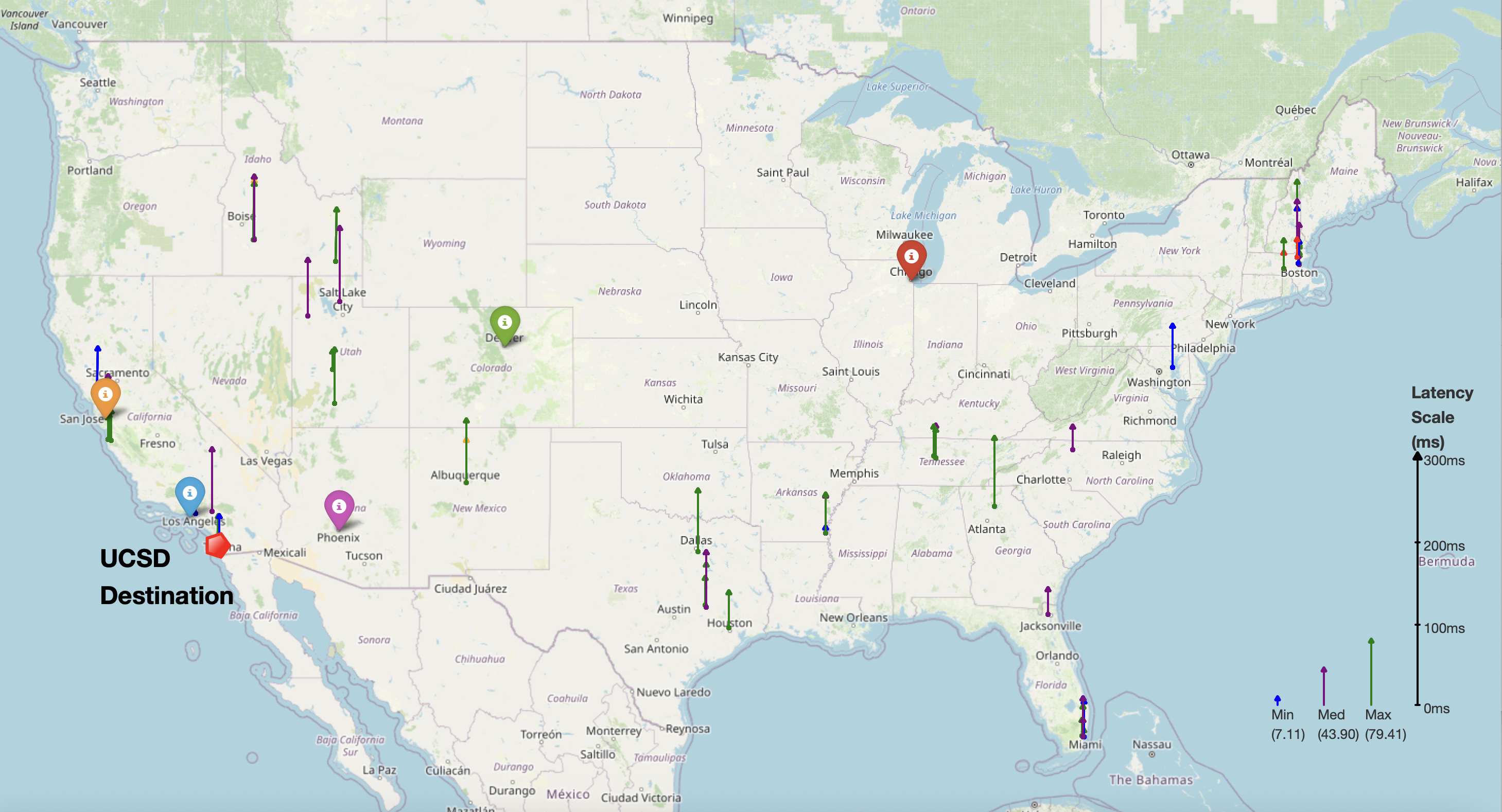}
    };
    \begin{scope}[x={(img.south east)}, y={(img.north west)}]

    \end{scope}
  \end{tikzpicture}
}\hfill
\subfloat[Verizon: 9 exits: Portland, Los Angeles, Dallas, Chicago, Cleveland, Richmond, Ashburn, NewYork, Boston; and add. direct path latency.]{
  \begin{tikzpicture}
    \node[anchor=south west, inner sep=0] (img) at (0,0) {
      \includegraphics[width=.31\textwidth]{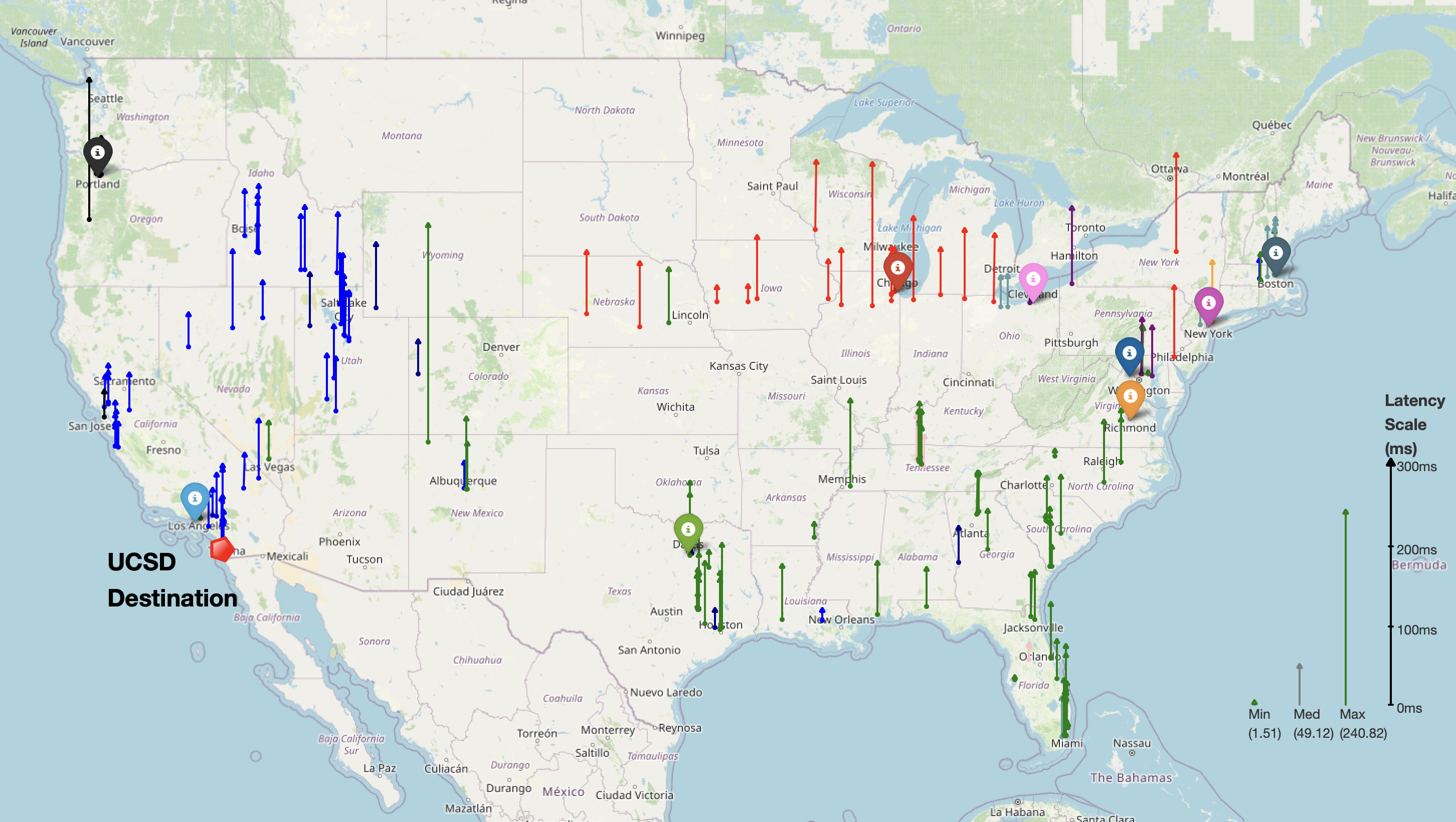}
    };
    \begin{scope}[x={(img.south east)}, y={(img.north west)}]

    \end{scope}
  \end{tikzpicture}
}

\caption{Observed additional direct-path latency (line height)
  and choice of exit point (line color),
  corresponding with observed exit points for each PLMN (inverted raindrop markers),
  inferred from traceroutes from mobile phones across the U.S.
  The solid lines in \autoref{fig:addition_latency_cdfs} give a CDF of additional direct-path latency.
}
\label{fig:exit_point_locations}
\end{figure*}

Each vertical line in \autoref{fig:exit_point_locations} represents a latency observation from a handset to San Diego at different times and across different MNOs. 
Line height corresponds to additional direct-path latency. 
The number of observations varies by MNO due to differing traceroute success rates, 
and we report only one traceroute per unique cellular tower location.

\begin{figure}
%
  \label{fig:additional_latency_cdf}
  \includegraphics[width=0.7\columnwidth]{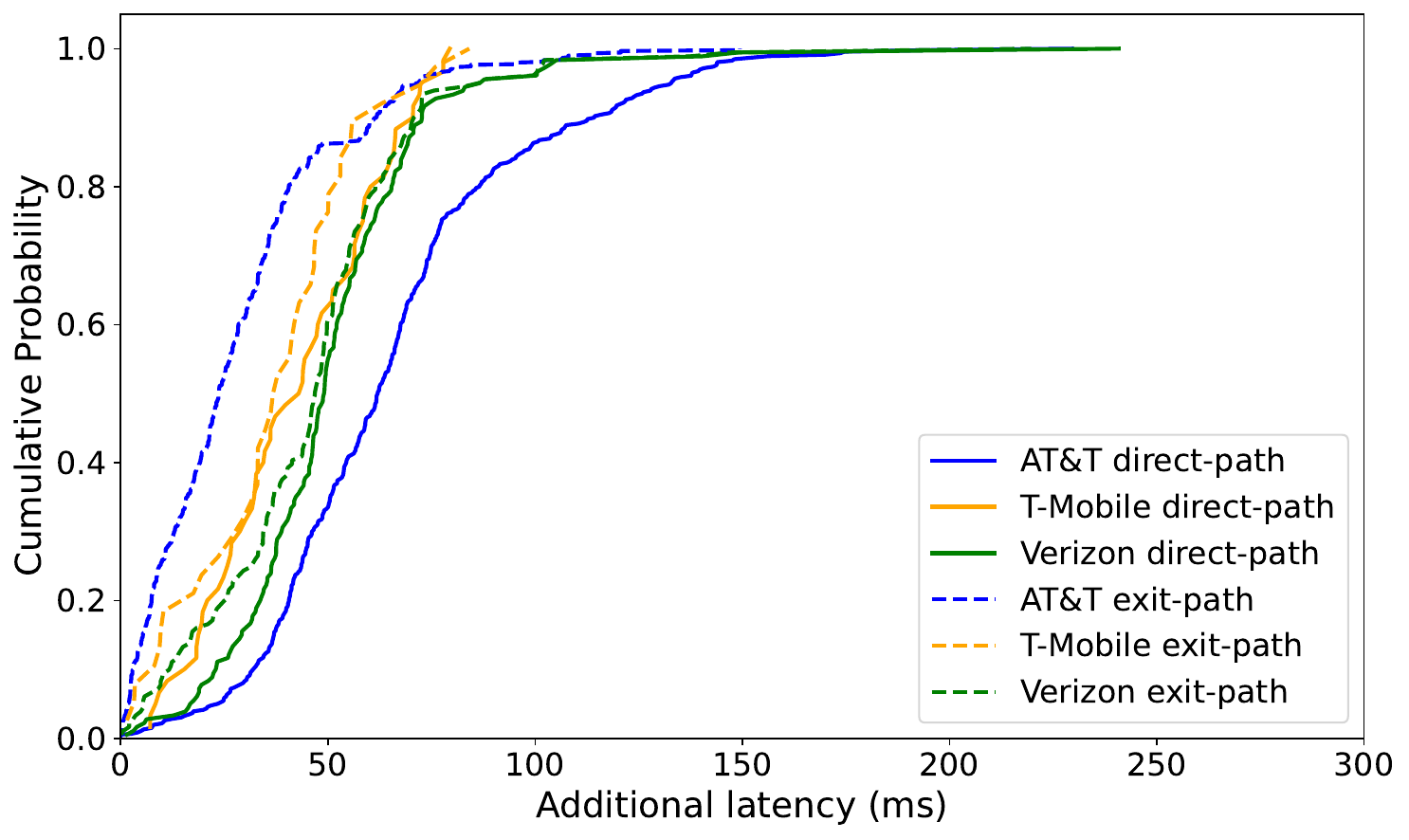}
\vspace{-7pt}
\caption{Additional direct-path (solid) and exit-path (dotted) latency per MNO (color).}
	\label{fig:addition_latency_cdfs}
\end{figure}

Latency is high even for mobile phones near the destination. The solid lines in \autoref{fig:addition_latency_cdfs} show the CDF of additional direct-path latency for each MNO, 
with median values of 62.1 ms for AT\&T, 43.9 ms for T-Mobile, and 49.1 ms for Verizon. 
For context, the speed-of-fiber latency from Los Angeles to Boston (4200 km) is 21 ms, 
while measured latencies are around 70 ms,
making an additional 40–60 ms significant.

\emph{Latency Causes:}
\autoref{fig:latency_methodology} shows additional direct-path latency for all observations passing through AT\&T's Chicago exit point. 
While \autoref{fig:exit_point_locations} shows the additional direct-path latency across all exit points for AT\&T, 
\autoref{fig:latency_methodology} zooms in on a single example: AT\&T’s Chicago exit point, to make the pattern of increasing latency with distance more visually salient.
In this focused view, we can observe that the latency is lowest when the handset is near Chicago, 
indicating that once traffic reaches the Internet, 
it follows a relatively direct path. 
However, 
latency increases as handsets move farther away, 
peaking for devices in North Dakota and New York due to traffic being relayed through Chicago, 
sometimes backtracking before reaching San Diego.

\emph{Hot-Potato routing:}
 Handsets in Nebraska and Kansas exit through Chicago
  rather than Los Angeles (another exit point of AT\&T),
  although Los Angeles lies on the path to San Diego~(\autoref{fig:latency_methodology}).
MNO sends traffic to the exit point nearest the handset,
  apparently without considering the location of the final destination
    in the Internet (\autoref{fig:exit_point_locations}).
This early-exit policy is an example of hot-potato routing,
  long known to inflate network latency~\cite{Spring23}.

Our latency observations are not affected by radio-specific latency.
Mobile phones have sophisticated algorithms to sleep the radio to conserve energy~\cite{narayanan2021variegated, rosen2014discovering}.
Moreover, 
ShipTraceroute initiates each traceroute after an airplane-mode toggle,
  so only the first probe can incur any radio startup delay.
Our observations use final RTT 
  which occurs after the connection is established and the radio is active,
  so it does not include the radio wake-up delay.

  We also confirm route stability over short timescales (hours).
The ShipTraceroute authors report no destination-dependent routing
  or traffic engineering, and no path change as the phone moves
  within a region \cite{Zhang21b}.
We find the same, with traceroutes from a given device and location
  always traverse the same exit point.

\subsection{Additional Exit-Path Latency}
	\label{sec:result_exit_latency}

Since the requirement to pass through an exit point is a large factor
  in mobile-to-Internet latency, 
  we next look at \emph{additional exit-path latency},
  the additional latency that occurs once we accept that we must use a given exit point.
Additional exit-path latency represents the cost of routing
  and indirect paths,
  but it factors out the physical latency incurred by passing through an exit point.

\begin{figure*}[t]
\centering
\subfloat[Additional exit-path latency of AT\&T]{
  \label{fig:additional_exit_latency_map_att}
  \begin{tikzpicture}
    \node[anchor=south west, inner sep=0] (img) at (0,0) {
      \includegraphics[width=.31\textwidth]{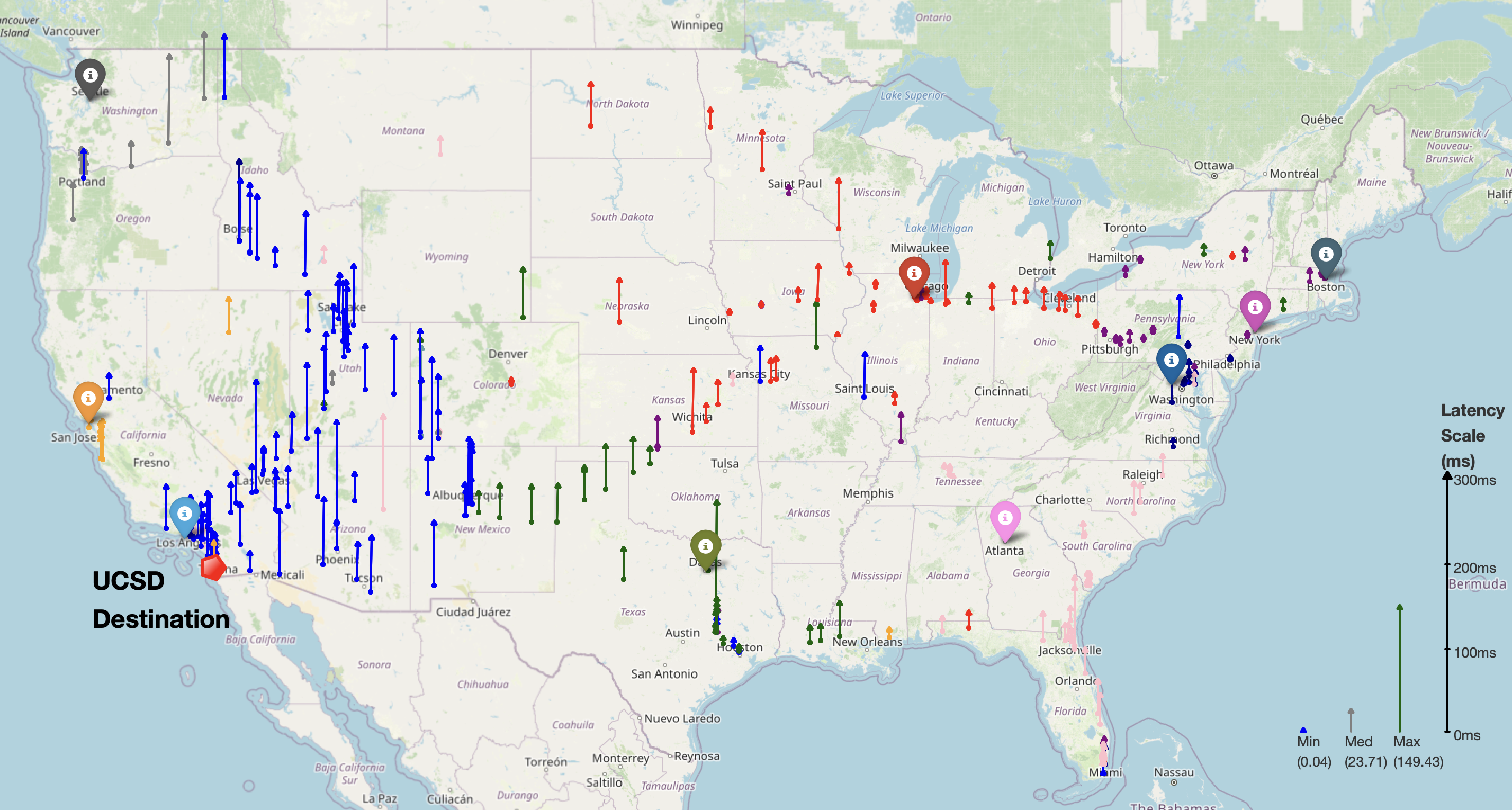}
    };
    \begin{scope}[x={(img.south east)}, y={(img.north west)}]
      \draw[thin, dashed] (0.38, 0.67) rectangle (0.50, 0.92);
      \node[font=\scriptsize\bfseries, align=center] at (0.44, 0.795) {Dakotas};

       \draw[thin, dashed] (0.22,0.33) rectangle (0.36,0.50);
\node[anchor=center, font=\scriptsize\bfseries, align=left] 
  at (0.27,0.405) {Four\\Corners};
    \end{scope}
  \end{tikzpicture}
}\hfill
\subfloat[Additional exit-path latency of T-Mobile]{
  \begin{tikzpicture}
    \node[anchor=south west, inner sep=0] (img) at (0,0) {
      \includegraphics[width=.30\textwidth]{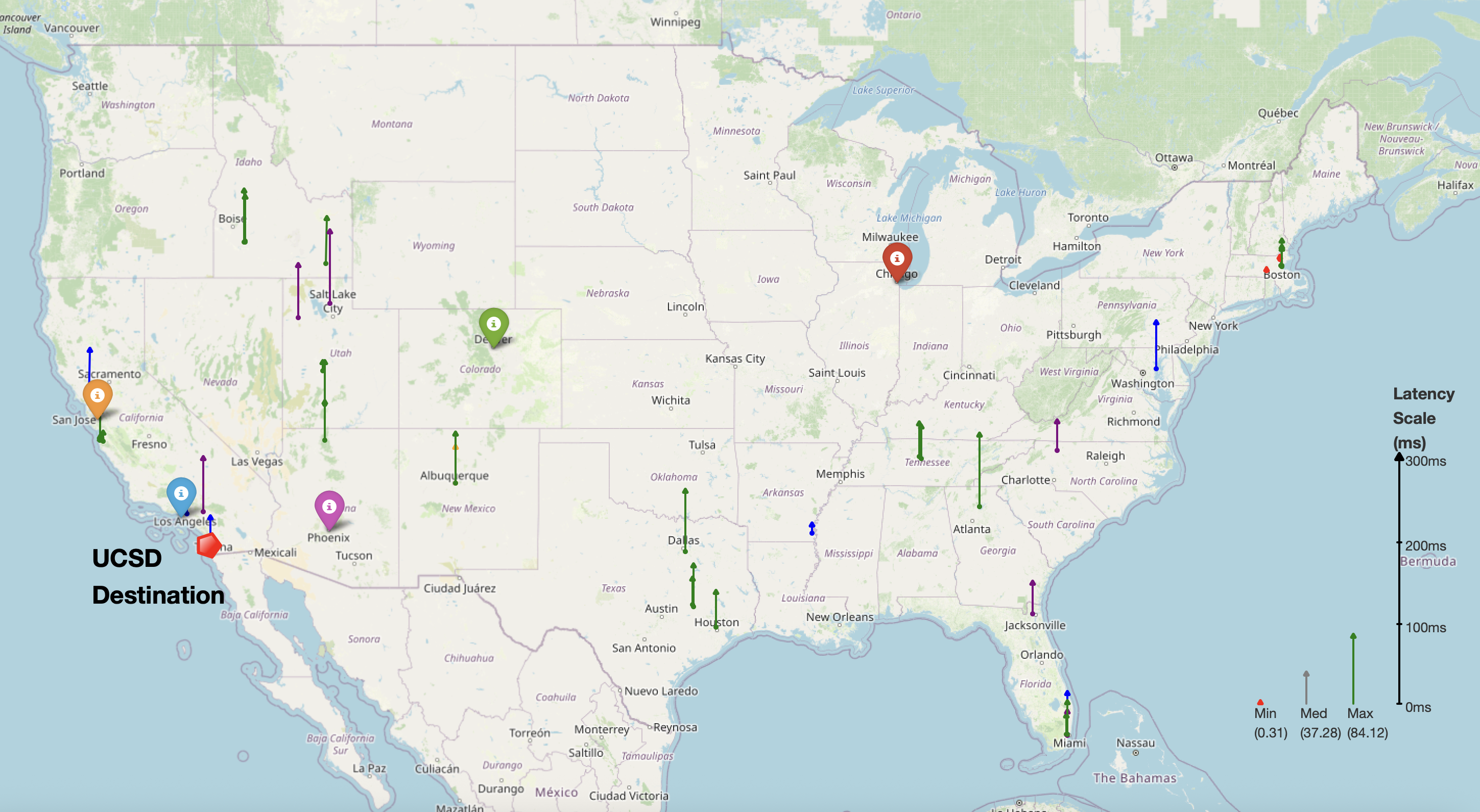}
    };
    \begin{scope}[x={(img.south east)}, y={(img.north west)}]
    \end{scope}
  \end{tikzpicture}
}\hfill
\subfloat[Additional exit-path latency of Verizon]{
  \begin{tikzpicture}
    \node[anchor=south west, inner sep=0] (img) at (0,0) {
      \includegraphics[width=.295\textwidth]{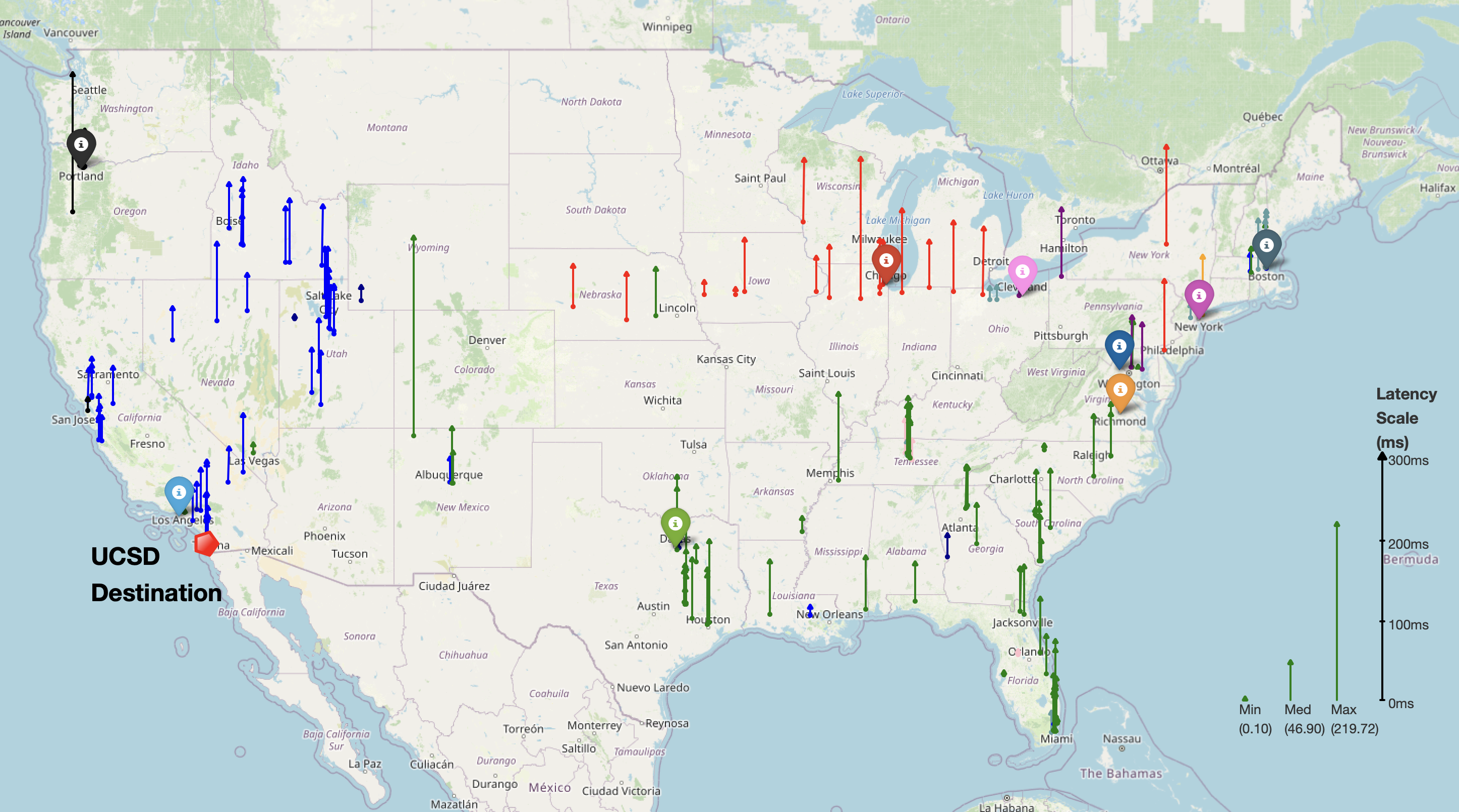}
    };
    \begin{scope}[x={(img.south east)}, y={(img.north west)}]
    \end{scope}
  \end{tikzpicture}
}
\vspace{-10pt}
\caption{Observed additional exit-path latency (line height) and exit-point selection (line color)
  for each MNO (inverted raindrop markers).
Dotted lines in \autoref{fig:addition_latency_cdfs} give the CDF.
  }
\label{fig:exit_path_latency_maps}
\end{figure*}

\autoref{fig:exit_path_latency_maps} shows additional exit-path latency
  for each MNO,
  with \autoref{fig:addition_latency_cdfs} showing corresponding CDFs.
The median additional exit-path latency is 23.6\,ms (AT\&T),
  37.2\,ms (T-Mobile), and 46.9\,ms (Verizon),
  all less than additional direct-path latency,
  reflecting the cost of relay through an exit point.
We assume the  wireless-hop latency is similar across providers,
  leaving a large 23\,ms difference between AT\&T and Verizon.
To better understand how geographic distance influences exit-path performance,
we analyze the relationship between round-trip path length and additional
exit-path latency.
We find a negative correlation of AT\&T $r=-0.342$,
T-Mobile
$r=-0.364$, 
and Verizon $r=-0.181$;
  with details
 in Appendix~\autoref{app:distance_latency}.
Given the extra traversal to the gateway, 
exit‐path latency always exceeds the direct path;
but our contribution is to quantify the margin of the cost.
For T-Mobile and Verizon, this reduction is minimal (4 to 7 ms), 
as the fiber optic speed allows 5 ms to cover approximately 1000\,km. In contrast, 
AT\&T's additional exit-path latency is 38.5 ms lower than its additional direct-path latency, 
suggesting significant routing inefficiencies, 
particularly in the Dakotas (North and South Dakota) and the Four Corners region (Regions marked with dashed boxes in \autoref{fig:exit_point_locations} and \autoref{fig:exit_path_latency_maps} for AT\&T.) 
This discrepancy may stem from greater data availability for AT\&T, as T-Mobile and Verizon have limited data for these regions.

\begin{figure*}[t]
\centering
\subfloat[AT\&T 9-exit optimal:
   Seattle, WA; Bakersfield, CA; 4-Corners, NM; Dallas, TX; Des Moines, IA; Gary, IN; Washington, DC; White Plains, NY;
 compared to known exits]{
    \label{fig:optimal_9_exit_points}
    \includegraphics[width=.29\textwidth]{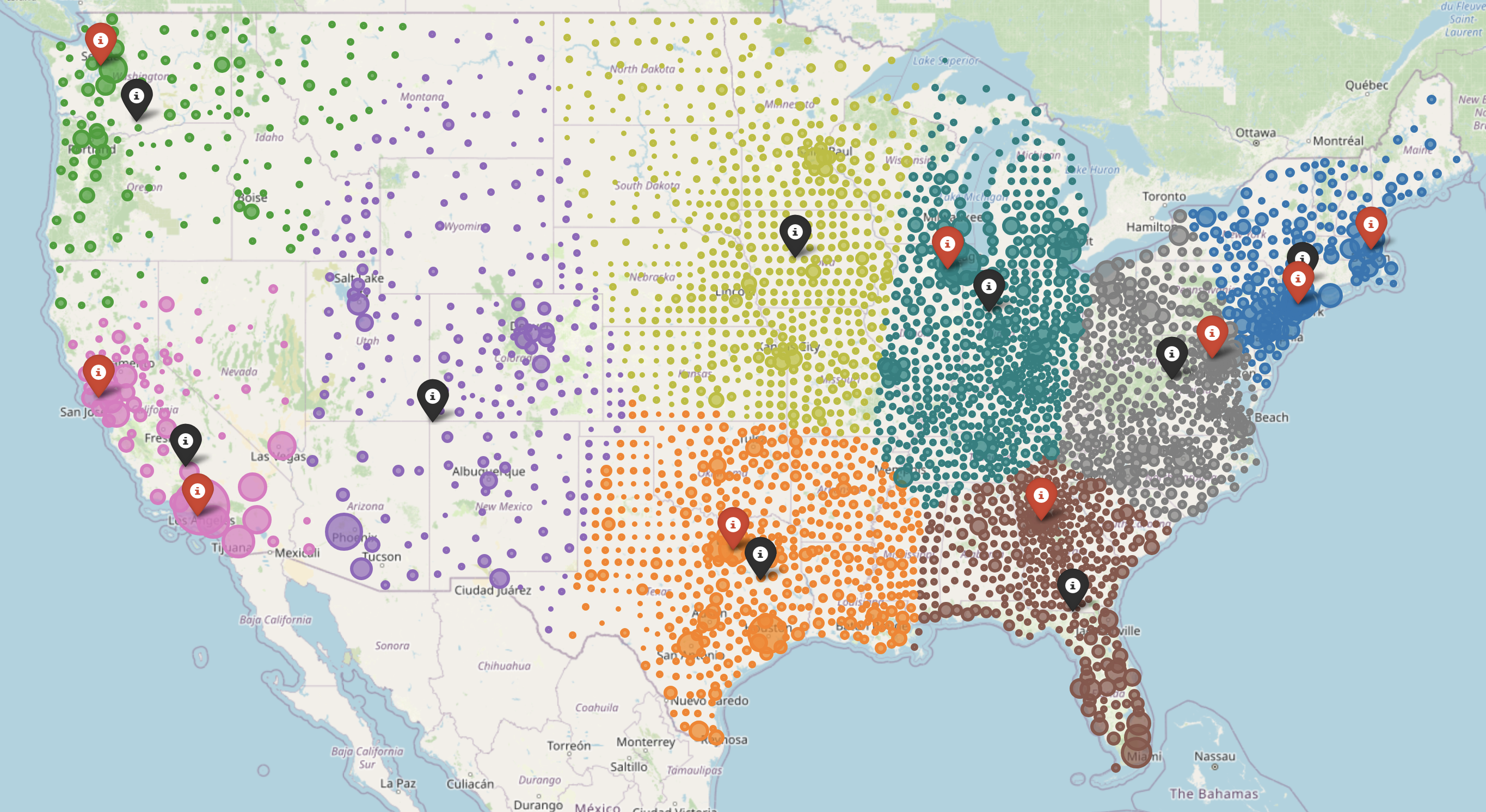}
}\hfill
\subfloat[T-Mobile 5-exit optimal:
Reno, NV; Gainesville, TX; Chicago, IL; Macon, GA; New York, NY;
 compared to known exits]{
    \label{fig:optimal_5_exit_points}
    \includegraphics[width=.30\textwidth]{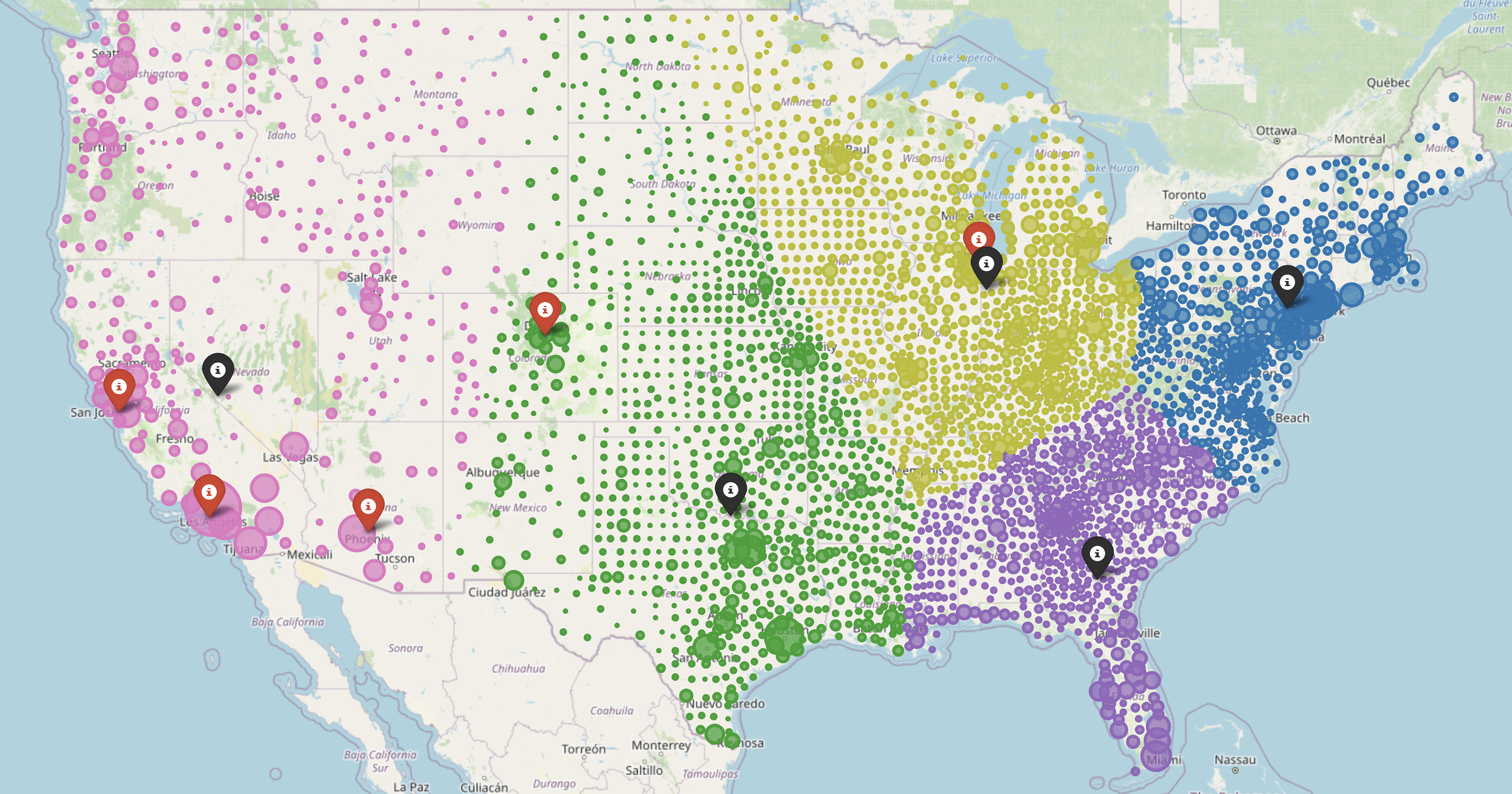}
}\hfill
\subfloat[Verizon 9-exit optimal:
   Seattle, WA; Bakersfield, CA; 4-Corners, NM; Dallas, TX; Des Moines, IA; Gary, IN; Washington, DC; White Plains, NY;
 compared to known exits]{
    \label{fig:Verizon_optimal_exitpoints}
    \includegraphics[width=.29\textwidth]{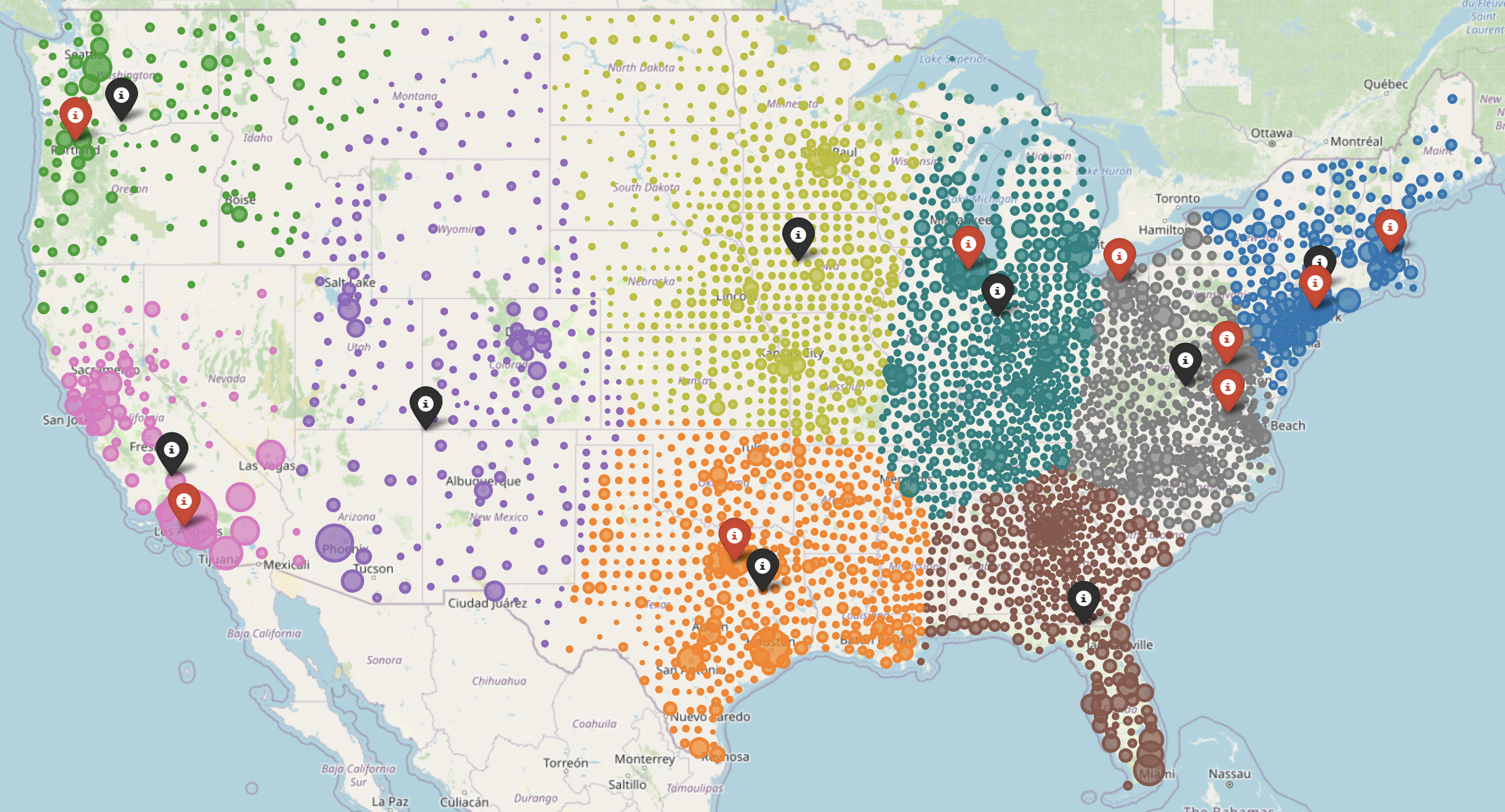}
}
\vspace{-7pt}
\caption{Theoretical optimal exit-point placement (black teardrops) and catchments (colored circles) for 9- and 5-exit-point networks, 
compared with known exit points (red teardrops), weighted by U.S. population (circle area).}
  \label{fig:optimal_exit_points}
\end{figure*}

\subsection{Optimal Exit Location and Number}
	\label{sec:exit_optimality}

\textbf{Optimal locations:}
We evaluate optimal locations for exit points
  based on U.S. population density (\autoref{sec:optimizing_location_methodology}).
\autoref{fig:optimal_exit_points} shows our
  computed optimal placements (black teardrops)
  and their corresponding catchments (colored circles)
  with both 9- and 5-exits.
We compare these optimal placements with the known exit points for AT\&T (\autoref{fig:optimal_9_exit_points}), 
T-Mobile (\autoref{fig:optimal_5_exit_points}), 
and Verizon (\autoref{fig:Verizon_optimal_exitpoints}).

Optimal exit placements reflect population density, 
with a higher concentration in eastern and mid-western regions of the U.S. 
However, the model does not account for existing telecommunications infrastructure. 
For example, it suggests an exit near the Four Corners region, despite limited local infrastructure, 
and places an exit in Bakersfield, California, 
rather than in metropolitan areas such as Los Angeles or San Francisco. 
Although Bakersfield appears to be an unusual choice, 
it is strategically located between major population centers and along key transit corridors. 
This placement could benefit rural users, those in areas with limited Wi-Fi access, and travelers across California by reducing latency.

A comparison of known and optimal placements shows that AT\&T and Verizon's exit points are well-aligned with the population distribution, especially in major metropolitan areas such as Seattle, Los Angeles, Dallas, Chicago, Atlanta, Washington, D.C., and New York. 
On the other hand, T-Mobile's five exits show larger variance from optimal,
  with two exit points in California and none on the East Coast. 

\textbf{Efficiency of actual placements:}
To quantify the alignment of known and optimal placements, we compute an \textit{exit-efficiency} ratio, 
which compares the per-capita distance to the nearest known versus optimal exit. 
A ratio of 1 indicates perfect alignment, 
while values above 1 indicate additional latency. AT\&T achieves an exit-efficiency ratio of 1.18, 
while Verizon and T-Mobile have ratios of 1.43 and 1.66, respectively. 
These values suggest that AT\&T's exit points are well-placed, Verizon's are moderately efficient, and T-Mobile's limited coverage contributes to higher latency, 
particularly in the East and South.

We also analyze exit-point placement
  with MNO customers.
We estimate customers by weighting population data
  by publicly reported MNO market share~\cite{marketshare}, 
Market-share weighted ratios are 1.32 for AT\&T, 
  1.72 for Verizon, 
  and 1.61 for T-Mobile.
These again suggest that AT\&T operates closer to population than other MNOs.
However, Verizon's market-share-weighted exit points are further from optimal than population-weighted and is comparable for T-Mobile.

Exit-point placement is inherently constrained by network topology, capacity limitations, 
the need to maintain balanced link utilization,
and related business constraints, such as high-demand areas and regional prioritization.
However, 
our analysis quantifies the latency cost of current deployment choices, providing a benchmark for potential optimizations. 
While operators cannot freely place exit points, 
our findings suggest that targeted adjustments %
could reduce latency without increasing operational complexity.

\textbf{Benefits of more:}
We also extend our analysis to ask:
\emph{How significant is the latency improvement achieved by deploying extra exit points?}
We expect additional exit points to lower mean latency per capita.
Our analysis shows that each new exit yields only modest improvement,
  reducing latency by 1\% or less
 (details in \autoref{app:optimizing_n+i_location}).
Optimizing the locations of existing exits provides a larger reduction in latency.
\emph{Better placement} is more beneficial than simply adding more exits.

\section{Validation}
\label{sec:validation}

We validate the stability of inferred exit points
  by comparing exit-point locations seen in 2021 and 2024.
For 2021, we use ShipTraceroute,
  and for 2024 use Northeastern's 5G driving data~\cite{ghoshal2025replication}.
The 2024 data is sparser than 2021,
  reporting 439 traceroutes for AT\&T and 441 for Verizon. 

Northeastern's data finds
  some of exit points we see in 2021:
  we see 3 of 9 exits (33.3\%, in San Francisco, L.A., and Ashburn) for AT\&T,
  and 5 of 9 (55.6\%, in L.A., Chicago, Ashburn, N.Y., and Boston) for Verizon.
It also finds a \emph{new} exit in Pittsburgh for Verizon.
The persistence of these locations across datasets indicates that exit-point placements are stable over years,
  as expected since an exit requires significant engineering to deploy.
While our results confirm many exit points are still active in the same locations
  (precision is high),
  sparseness in the Northeastern data
  prevents assessing completeness.
We cannot validate T-Mobile exits because
  policy changes hide traceroute after 2021.


\section{Conclusion}

We showed how to identify components of latency
  in how MNO networks interact with the Internet.
We show that exit-point interconnections add latency,
  but relatively few exit points can provide good latency for the U.S. 

\section*{Acknowledgments}
We thank the authors of ShipTraceroute and the Northeastern (cellular-network-performance-on-wheels) dataset for their timely responses and clarification about the datasets.

This material is based upon work supported by DARPA under Contract No.~HR001120C0157 (SABRES),
  and by NSF under OAC-2530698 (BRIPOD) and 2453092 (TEPPIT).
Any opinions, findings and conclusions or recommendations expressed in this material are those of the author(s) and do not necessarily reflect the views of DARPA or NSF.

\bibliographystyle{ACM-Reference-Format}

\begin{thebibliography}{62}


\ifx \showCODEN    \undefined \def \showCODEN     #1{\unskip}     \fi
\ifx \showDOI      \undefined \def \showDOI       #1{#1}\fi
\ifx \showISBNx    \undefined \def \showISBNx     #1{\unskip}     \fi
\ifx \showISBNxiii \undefined \def \showISBNxiii  #1{\unskip}     \fi
\ifx \showISSN     \undefined \def \showISSN      #1{\unskip}     \fi
\ifx \showLCCN     \undefined \def \showLCCN      #1{\unskip}     \fi
\ifx \shownote     \undefined \def \shownote      #1{#1}          \fi
\ifx \showarticletitle \undefined \def \showarticletitle #1{#1}   \fi
\ifx \showURL      \undefined \def \showURL       {\relax}        \fi
%
%
\providecommand\bibfield[2]{#2}
\providecommand\bibinfo[2]{#2}
\providecommand\natexlab[1]{#1}
\providecommand\showeprint[2][]{arXiv:#2}

\bibitem[mar({[n.\,d.]})]%
        {marketshare}
 \bibinfo{year}{[n.\,d.]}\natexlab{}.
\newblock \bibinfo{title}{{Top Mobile Data Providers State-by-State}}.
\newblock
  \bibinfo{howpublished}{\url{https://www.webfx.com/blog/internet/top-mobile-data-providers-state-by-state/}}.
\newblock


\bibitem[Ozcan et~al\mbox{.}(2023)]%
        {ozcan23}
\bibfield{author}{\bibinfo{person}{Selim Ozcan},
  \bibinfo{person}{Ioana Livadariu},
  \bibinfo{person}{Georgios Smaragdakis}, {and}
  \bibinfo{person}{Carsten Griwodz}.}
  \bibinfo{year}{2023}\natexlab{}.
\newblock \showarticletitle{Longitudinal Analysis of Inter-City
  Network Delays}. In
  \bibinfo{booktitle}{\emph{2023 7th Network Traffic Measurement
  and Analysis Conference (TMA)}}. IEEE.
\newblock
\urldef\tempurl%
\url{https://doi.org/10.23919/TMA58422.2023.10198987}
\showDOI{\tempurl}


\bibitem[{Analysis of Network Traffic (ANT) group}(2023)]%
        {ant_block_to_as}
\bibfield{author}{\bibinfo{person}{{Analysis of Network Traffic (ANT) group}}.}
  \bibinfo{year}{2023}\natexlab{}.
\newblock \bibinfo{title}{{LANDER:internet outage adaptive a52
  supplement-20230401}}.
\newblock
  \bibinfo{howpublished}{\url{https://ant.isi.edu/datasets/readmes/internet_outage_adaptive_a52_supplement-20230401.README.txt}}.
\newblock
\newblock
\shownote{[Online; accessed October-2023]}.


\bibitem[Behravesh et~al\mbox{.}(2024)]%
        {behravesh24}
\bibfield{author}{\bibinfo{person}{Rasoul Behravesh}, \bibinfo{person}{David
  Breitgand}, \bibinfo{person}{Dean~H Lorenz}, {and} \bibinfo{person}{Danny
  Raz}.} \bibinfo{year}{2024}\natexlab{}.
\newblock \showarticletitle{{A practical near optimal deployment of service
  function chains in edge-to-cloud networks}}. In
  \bibinfo{booktitle}{\emph{IEEE INFOCOM}}. IEEE, \bibinfo{pages}{751--760}.
\newblock


\bibitem[Bradley et~al\mbox{.}(2000)]%
        {Bradley00a}
\bibfield{author}{\bibinfo{person}{P.~S. Bradley}, \bibinfo{person}{K.~P.
  Bennett}, {and} \bibinfo{person}{A. Demiriz}.}
  \bibinfo{year}{2000}\natexlab{}.
\newblock \bibinfo{booktitle}{\emph{Constrained K-Means Clustering}}.
\newblock \bibinfo{type}{{T}echnical {R}eport} MSR-TR-2000-65.
  \bibinfo{institution}{Microsoft Research}.
\newblock
\urldef\tempurl%
\url{https://www.microsoft.com/en-us/research/wp-content/uploads/2016/02/tr-2000-65.pdf}
\showURL{%
\tempurl}


\bibitem[Brakmo and Peterson(1995)]%
        {Brakmo95a}
\bibfield{author}{\bibinfo{person}{L.S. Brakmo} {and} \bibinfo{person}{L.L.
  Peterson}.} \bibinfo{year}{1995}\natexlab{}.
\newblock \showarticletitle{{TCP Vegas: end to end congestion avoidance on a
  global Internet}}.
\newblock \bibinfo{journal}{\emph{IEEE Journal on Selected Areas in
  Communications}} \bibinfo{volume}{13}, \bibinfo{number}{8}
  (\bibinfo{year}{1995}), \bibinfo{pages}{1465--1480}.
\newblock
\urldef\tempurl%
\url{https://doi.org/10.1109/49.464716}
\showDOI{\tempurl}


\bibitem[Calder et~al\mbox{.}(2013)]%
        {Calder13a}
\bibfield{author}{\bibinfo{person}{Matt Calder}, \bibinfo{person}{Xun Fan},
  \bibinfo{person}{Zi Hu}, \bibinfo{person}{Ethan Katz-Bassett},
  \bibinfo{person}{John Heidemann}, {and} \bibinfo{person}{Ramesh Govindan}.}
  \bibinfo{year}{2013}\natexlab{}.
\newblock \showarticletitle{{Mapping the Expansion of {Google's} Serving
  Infrastructure}}. In \bibinfo{booktitle}{\emph{Proceedings of the ACM
  Internet Measurement Conference}}. \bibinfo{publisher}{{ACM}},
  \bibinfo{address}{Barcelona, Spain}, \bibinfo{pages}{313--326}.
\newblock
\urldef\tempurl%
\url{https://doi.org/10.1145/2504730.2504754}
\showDOI{\tempurl}


\bibitem[Cornu{\'e}jols et~al\mbox{.}(1983)]%
        {uncapicitated83}
\bibfield{author}{\bibinfo{person}{G{\'e}rard Cornu{\'e}jols},
  \bibinfo{person}{George Nemhauser}, {and} \bibinfo{person}{Laurence Wolsey}.}
  \bibinfo{year}{1983}\natexlab{}.
\newblock \bibinfo{booktitle}{\emph{{The uncapacitated facility location
  problem}}}.
\newblock \bibinfo{type}{{T}echnical {R}eport}. \bibinfo{institution}{Cornell
  University Operations Research and Industrial Engineering}.
\newblock


\bibitem[Deng et~al\mbox{.}(2018)]%
        {deng2018mobility}
\bibfield{author}{\bibinfo{person}{Haotian Deng}, \bibinfo{person}{Chunyi
  Peng}, \bibinfo{person}{Ans Fida}, \bibinfo{person}{Jiayi Meng}, {and}
  \bibinfo{person}{Y~Charlie Hu}.} \bibinfo{year}{2018}\natexlab{}.
\newblock \showarticletitle{{Mobility support in cellular networks: A
  measurement study on its configurations and implications}}. In
  \bibinfo{booktitle}{\emph{Proc. of the Internet Measurement Conference}}.
\newblock


\bibitem[Durairajan et~al\mbox{.}(2015)]%
        {Durairajan15a}
\bibfield{author}{\bibinfo{person}{Ramakrishnan Durairajan},
  \bibinfo{person}{Paul Barford}, \bibinfo{person}{Joel Sommers}, {and}
  \bibinfo{person}{Walter Willinger}.} \bibinfo{year}{2015}\natexlab{}.
\newblock \showarticletitle{{InterTubes}: A Study of the {US} Long-haul
  Fiber-optic Infrastructure}. In \bibinfo{booktitle}{\emph{ACM SIGCOMM}}.
  \bibinfo{publisher}{{ACM}}, \bibinfo{address}{London, United Kingdom},
  \bibinfo{pages}{565--578}.
\newblock
\urldef\tempurl%
\url{https://doi.org/10.1145/2785956.2787499}
\showDOI{\tempurl}


\bibitem[Eiselt and Marianov(2011)]%
        {eiselt2011pioneering}
\bibfield{author}{\bibinfo{person}{H.~A. Eiselt} {and}
  \bibinfo{person}{Vladimir Marianov}.} \bibinfo{year}{2011}\natexlab{}.
\newblock \showarticletitle{{Pioneering Developments in Location Analysis:
  Location Problems—Its Problem Statement, Its Components, and Application}}.
\newblock In \bibinfo{booktitle}{\emph{Foundations of Location Analysis}},
  \bibfield{editor}{\bibinfo{person}{H.~A. Eiselt} {and}
  \bibinfo{person}{Vladimir Marianov}} (Eds.). \bibinfo{publisher}{Springer},
  \bibinfo{pages}{3--6}.
\newblock
\showISBNx{978-1-4419-7571-3}


\bibitem[Equinix({[n.\,d.]})]%
        {Datacenters}
\bibfield{author}{\bibinfo{person}{Equinix}.}
  \bibinfo{year}{[n.\,d.]}\natexlab{}.
\newblock \bibinfo{title}{{Equinix America data centers}}.
\newblock
  \bibinfo{howpublished}{\url{https://www.equinix.com/data-centers/americas-colocation}}.
\newblock


\bibitem[{Feng, Yingying and Zhao, Shasha and Liu, Hui}(2020)]%
        {Feng20}
\bibfield{author}{\bibinfo{person}{{Feng, Yingying and Zhao, Shasha and Liu,
  Hui}}.} \bibinfo{year}{2020}\natexlab{}.
\newblock \showarticletitle{Analysis of Network Coverage Optimization Based on
  Feedback K-Means Clustering and Artificial Fish Swarm Algorithm}.
\newblock \bibinfo{journal}{\emph{IEEE Access}}  \bibinfo{volume}{8}
  (\bibinfo{year}{2020}).
\newblock
\urldef\tempurl%
\url{https://doi.org/10.1109/ACCESS.2020.2970208}
\showDOI{\tempurl}


\bibitem[Frangoudis et~al\mbox{.}(2017)]%
        {frangoudis2017cdn}
\bibfield{author}{\bibinfo{person}{Pantelis~A Frangoudis},
  \bibinfo{person}{Louiza Yala}, {and} \bibinfo{person}{Adlen Ksentini}.}
  \bibinfo{year}{2017}\natexlab{}.
\newblock \showarticletitle{{CDN-as-a-service provision over a telecom
  operator’s cloud}}.
\newblock \bibinfo{journal}{\emph{IEEE Transactions on Network and Service
  Management}} \bibinfo{volume}{14}, \bibinfo{number}{3}
  (\bibinfo{year}{2017}), \bibinfo{pages}{702--716}.
\newblock


\bibitem[Gharaibeh et~al\mbox{.}(2017)]%
        {Gharaibeh17}
\bibfield{author}{\bibinfo{person}{Manaf Gharaibeh}, \bibinfo{person}{Anant
  Shah}, \bibinfo{person}{Bradley Huffaker}, \bibinfo{person}{Han Zhang},
  \bibinfo{person}{Roya Ensafi}, {and} \bibinfo{person}{Christos
  Papadopoulos}.} \bibinfo{year}{2017}\natexlab{}.
\newblock \showarticletitle{A look at router geolocation in public and
  commercial databases}. In \bibinfo{booktitle}{\emph{Proceedings of the 2017
  Internet Measurement Conference}} (London, United Kingdom)
  \emph{(\bibinfo{series}{IMC '17})}. \bibinfo{publisher}{ACM},
  \bibinfo{address}{New York, NY, USA}, \bibinfo{pages}{463–469}.
\newblock
\showISBNx{9781450351188}
\urldef\tempurl%
\url{https://doi.org/10.1145/3131365.3131380}
\showDOI{\tempurl}


\bibitem[Ghoshal et~al\mbox{.}({[n.\,d.]})]%
        {ghoshal2025replication}
\bibfield{author}{\bibinfo{person}{Moinak Ghoshal}, \bibinfo{person}{Omar
  Basit}, \bibinfo{person}{Imran Khan}, \bibinfo{person}{Z~Jonny Kong},
  \bibinfo{person}{Sizhe Wang}, \bibinfo{person}{Yufei Feng},
  \bibinfo{person}{Phuc Dinh}, \bibinfo{person}{Y~Charlie Hu}, {and}
  \bibinfo{person}{Dimitrios Koutsonikolas}.}
  \bibinfo{year}{[n.\,d.]}\natexlab{}.
\newblock \showarticletitle{{Replication: Performance of Cellular Networks on
  the Wheels}}. In \bibinfo{booktitle}{\emph{Proc. of the 2025 ACM IMC}}.
\newblock


\bibitem[Ghoshal et~al\mbox{.}(2023)]%
        {Ghoshal23}
\bibfield{author}{\bibinfo{person}{Moinak Ghoshal}, \bibinfo{person}{Imran
  Khan}, \bibinfo{person}{Z.~Jonny Kong}, \bibinfo{person}{Phuc Dinh},
  \bibinfo{person}{Jiayi Meng}, \bibinfo{person}{Y.~Charlie Hu}, {and}
  \bibinfo{person}{Dimitrios Koutsonikolas}.} \bibinfo{year}{2023}\natexlab{}.
\newblock \showarticletitle{{Performance of Cellular Networks on the Wheels}}.
  In \bibinfo{booktitle}{\emph{Proc. of the 2023 ACM on Internet Measurement
  Conference}} (Montreal QC, Canada) \emph{(\bibinfo{series}{IMC '23})}.
  \bibinfo{publisher}{ACM}, \bibinfo{address}{New York, NY, USA},
  \bibinfo{pages}{678–695}.
\newblock
\showISBNx{9798400703829}
\urldef\tempurl%
\url{https://doi.org/10.1145/3618257.3624814}
\showDOI{\tempurl}


\bibitem[Ghoshal et~al\mbox{.}(2025)]%
        {Ghoshal25b}
\bibfield{author}{\bibinfo{person}{Moinak Ghoshal}, \bibinfo{person}{Imran
  Khan}, \bibinfo{person}{Z.~Jonny Kong}, \bibinfo{person}{Yufei Feng},
  \bibinfo{person}{Phuc Dinh}, \bibinfo{person}{Jiayi Meng},
  \bibinfo{person}{Y.~Charlie Hu}, {and} \bibinfo{person}{Dimitrios
  Koutsonikolas}.} \bibinfo{year}{2025}\natexlab{}.
\newblock \bibinfo{title}{Data from the paper ``Performance of Cellular
  Networks on the Wheels''}.
\newblock \bibinfo{howpublished}{github repository
  \url{https://github.com/NUWiNS/imc2023-cellular-network-performance-on-wheels-data}}.
\newblock
\urldef\tempurl%
\url{https://github.com/NUWiNS/imc2023-cellular-network-performance-on-wheels-data}
\showURL{%
\tempurl}


\bibitem[Ghoshal et~al\mbox{.}(2022)]%
        {ghoshal2022depth}
\bibfield{author}{\bibinfo{person}{Moinak Ghoshal}, \bibinfo{person}{Z~Jonny
  Kong}, \bibinfo{person}{Qiang Xu}, \bibinfo{person}{Zixiao Lu},
  \bibinfo{person}{Shivang Aggarwal}, \bibinfo{person}{Imran Khan},
  \bibinfo{person}{Yuanjie Li}, \bibinfo{person}{Y~Charlie Hu}, {and}
  \bibinfo{person}{Dimitrios Koutsonikolas}.} \bibinfo{year}{2022}\natexlab{}.
\newblock \showarticletitle{{An in-depth study of uplink performance of 5G
  mmWave networks}}. In \bibinfo{booktitle}{\emph{Proc. of the ACM SIGCOMM
  Workshop on 5G and Beyond Network Measurements, Modeling, Use Cases}}.
\newblock


\bibitem[Gigis et~al\mbox{.}(2021)]%
        {Gigis21a}
\bibfield{author}{\bibinfo{person}{Petros Gigis}, \bibinfo{person}{Matt
  Calder}, \bibinfo{person}{Lefteris Manassakis}, \bibinfo{person}{George
  Nomikos}, \bibinfo{person}{Vasileios Kotronis}, \bibinfo{person}{Xenofontas
  Dimitropoulos}, \bibinfo{person}{Ethan Katz-Bassett}, {and}
  \bibinfo{person}{Georgios Smaragdakis}.} \bibinfo{year}{2021}\natexlab{}.
\newblock \showarticletitle{{Seven Years in the Life of Hypergiants'
  Off-Nets}}. In \bibinfo{booktitle}{\emph{Proceedings of the {ACM} SIGCOMM
  Conference}}. \bibinfo{publisher}{{ACM}}, \bibinfo{address}{Virtual},
  \bibinfo{pages}{516--534}.
\newblock
\urldef\tempurl%
\url{https://doi.org/10.1145/3452296.3472928}
\showDOI{\tempurl}


\bibitem[Google({[n.\,d.]})]%
        {GoogleDatacenters}
\bibfield{author}{\bibinfo{person}{Google}.}
  \bibinfo{year}{[n.\,d.]}\natexlab{}.
\newblock \bibinfo{title}{{Data Center locations published by Google}}.
\newblock
  \bibinfo{howpublished}{\url{https://www.google.com/about/datacenters/locations/}}.
\newblock


\bibitem[Gov({[n.\,d.]})]%
        {PopulationDensity}
\bibfield{author}{\bibinfo{person}{US Gov}.}
  \bibinfo{year}{[n.\,d.]}\natexlab{}.
\newblock \bibinfo{title}{{Population Density by county}}.
\newblock
  \bibinfo{howpublished}{\url{https://www.census.gov/data/datasets/time-series/demo/popest/2020s-counties-total.html}}.
\newblock


\bibitem[Hakimi(1964)]%
        {FacilityProblem}
\bibfield{author}{\bibinfo{person}{S.~Louis Hakimi}.}
  \bibinfo{year}{1964}\natexlab{}.
\newblock \showarticletitle{{Optimum Locations of Switching Centers and the
  Absolute Centers and Medians of a Graph}}.
\newblock \bibinfo{journal}{\emph{OR}}  \bibinfo{volume}{12}
  (\bibinfo{year}{1964}).
\newblock


\bibitem[He et~al\mbox{.}(2007)]%
        {Bing07}
\bibfield{author}{\bibinfo{person}{Bing He}, \bibinfo{person}{Bin Xie}, {and}
  \bibinfo{person}{Dharma~P. Agrawal}.} \bibinfo{year}{2007}\natexlab{}.
\newblock \showarticletitle{{Optimizing the Internet Gateway Deployment in a
  Wireless Mesh Network}}. In \bibinfo{booktitle}{\emph{2007 IEEE International
  Conference on Mobile Adhoc and Sensor Systems}}. \bibinfo{pages}{1--9}.
\newblock
\urldef\tempurl%
\url{https://doi.org/10.1109/MOBHOC.2007.4428603}
\showDOI{\tempurl}


\bibitem[IP2Location({[n.\,d.]})]%
        {IP2LOCATION}
\bibfield{author}{\bibinfo{person}{IP2Location}.}
  \bibinfo{year}{[n.\,d.]}\natexlab{}.
\newblock \bibinfo{title}{IP2Location Geolocation Lite}.
\newblock
  \bibinfo{howpublished}{\url{https://lite.ip2location.com/ip2location-lite}}.
\newblock


\bibitem[Jesus and Aguiar({[n.\,d.]})]%
        {jesus2012figures}
\bibfield{author}{\bibinfo{person}{Vitor Jesus} {and} \bibinfo{person}{Rui~L
  Aguiar}.} \bibinfo{year}{[n.\,d.]}\natexlab{}.
\newblock \showarticletitle{{Figures of merit for the placement (in) efficiency
  of interconnected CDNs}}. In \bibinfo{booktitle}{\emph{2012 IEEE Symposium on
  Computers and Communications (ISCC)}}. IEEE, \bibinfo{pages}{000277--000282}.
\newblock


\bibitem[K\"{a}m\"{a}r\"{a}inen et~al\mbox{.}(2017)]%
        {Kaemaeraeinen17}
\bibfield{author}{\bibinfo{person}{Teemu K\"{a}m\"{a}r\"{a}inen},
  \bibinfo{person}{Matti Siekkinen}, \bibinfo{person}{Antti
  Yl\"{a}-J\"{a}\"{a}ski}, \bibinfo{person}{Wenxiao Zhang}, {and}
  \bibinfo{person}{Pan Hui}.} \bibinfo{year}{2017}\natexlab{}.
\newblock \showarticletitle{{Dissecting the End-to-end Latency of Interactive
  Mobile Video Applications}}. In \bibinfo{booktitle}{\emph{Proceedings of the
  18th International Workshop on Mobile Computing Systems and Applications}}
  (Sonoma, CA, USA) \emph{(\bibinfo{series}{HotMobile '17})}.
  \bibinfo{publisher}{Association for Computing Machinery},
  \bibinfo{address}{New York, NY, USA}, \bibinfo{pages}{61–66}.
\newblock
\showISBNx{9781450349079}
\urldef\tempurl%
\url{https://doi.org/10.1145/3032970.3032985}
\showDOI{\tempurl}


\bibitem[Khan et~al\mbox{.}(2025)]%
        {khan25}
\bibfield{author}{\bibinfo{person}{Imran Khan}, \bibinfo{person}{Moinak
  Ghoshal}, \bibinfo{person}{Joana Angjo}, \bibinfo{person}{Sigrid Dimce},
  \bibinfo{person}{Mushahid Hussain}, \bibinfo{person}{Paniz Parastar},
  \bibinfo{person}{Yenchia Yu}, \bibinfo{person}{Xueting Deng},
  \bibinfo{person}{Sumit Hawal}, \bibinfo{person}{Shirui Huang},
  {et~al\mbox{.}}} \bibinfo{year}{2025}\natexlab{}.
\newblock \showarticletitle{{How mature is 5G deployment? A cross-sectional,
  year-long study of 5G uplink performance}}.
\newblock \bibinfo{journal}{\emph{Computer Communications}}
  \bibinfo{volume}{237} (\bibinfo{year}{2025}), \bibinfo{pages}{108153}.
\newblock


\bibitem[Koch et~al\mbox{.}(2021)]%
        {Koch21}
\bibfield{author}{\bibinfo{person}{Thomas Koch}, \bibinfo{person}{Ethan
  Katz-Bassett}, \bibinfo{person}{John Heidemann}, \bibinfo{person}{Matt
  Calder}, \bibinfo{person}{Calvin Ardi}, {and} \bibinfo{person}{Ke Li}.}
  \bibinfo{year}{2021}\natexlab{}.
\newblock \showarticletitle{{Anycast In context: a tale of two systems}}. In
  \bibinfo{booktitle}{\emph{Proceedings of the 2021 ACM SIGCOMM 2021
  Conference}} (Virtual Event, USA) \emph{(\bibinfo{series}{SIGCOMM '21})}.
  \bibinfo{publisher}{Association for Computing Machinery},
  \bibinfo{address}{New York, NY, USA}, \bibinfo{pages}{398–417}.
\newblock
\showISBNx{9781450383837}
\urldef\tempurl%
\url{https://doi.org/10.1145/3452296.3472891}
\showDOI{\tempurl}


\bibitem[Kousias et~al\mbox{.}(2022a)]%
        {kousias2022coverage}
\bibfield{author}{\bibinfo{person}{Konstantinos Kousias},
  \bibinfo{person}{Mohammad Rajiullah}, \bibinfo{person}{Giuseppe Caso},
  \bibinfo{person}{Ozgu Alay}, \bibinfo{person}{Anna Brunstrom},
  \bibinfo{person}{Luca De~Nardis}, \bibinfo{person}{Marco Neri},
  \bibinfo{person}{Usman Ali}, {and} \bibinfo{person}{Maria-Gabriella
  Di~Benedetto}.} \bibinfo{year}{2022}\natexlab{a}.
\newblock \showarticletitle{{Coverage and performance analysis of 5G
  non-standalone deployments}}. In \bibinfo{booktitle}{\emph{Proceedings of the
  16th ACM Workshop on Wireless Network Testbeds, Experimental evaluation \&
  CHaracterization}}. \bibinfo{pages}{61--68}.
\newblock


\bibitem[Kousias et~al\mbox{.}(2022b)]%
        {kousias2022implications}
\bibfield{author}{\bibinfo{person}{Konstantinos Kousias},
  \bibinfo{person}{Mohammad Rajiullah}, \bibinfo{person}{Giuseppe Caso},
  \bibinfo{person}{Ozgu Alay}, \bibinfo{person}{Anna Brunstrom},
  \bibinfo{person}{Luca De~Nardis}, \bibinfo{person}{Marco Neri},
  \bibinfo{person}{Usman Ali}, {and} \bibinfo{person}{Maria-Gabriella
  Di~Benedetto}.} \bibinfo{year}{2022}\natexlab{b}.
\newblock \showarticletitle{{Implications of handover events in commercial 5G
  non-standalone deployments in Rome}}. In
  \bibinfo{booktitle}{\emph{Proceedings of the ACM SIGCOMM Workshop on 5G and
  Beyond Network Measurements, Modeling, and Use Cases}}.
  \bibinfo{pages}{22--27}.
\newblock


\bibitem[Kulkarni and Heidemann(2026)]%
        {Kulkarni26b}
\bibfield{author}{\bibinfo{person}{Shefali~Tushar Kulkarni} {and}
  \bibinfo{person}{John Heidemann}.} \bibinfo{year}{2026}\natexlab{}.
\newblock \bibinfo{booktitle}{\emph{Bounding Mobile Latency from Network
  Interconnections}}.
\newblock \bibinfo{type}{{T}echnical {R}eport} arXiv:tbd [cs.NI].
  \bibinfo{institution}{USC/ISI}.
\newblock
\urldef\tempurl%
\url{https://ant.isi.edu/%7ejohnh/PAPERS/Kulkarni26b.html}
\showURL{%
\tempurl}


\bibitem[Li et~al\mbox{.}(2018)]%
        {Li18a}
\bibfield{author}{\bibinfo{person}{Zhihao Li}, \bibinfo{person}{Dave Levin},
  \bibinfo{person}{Neil Spring}, {and} \bibinfo{person}{Bobby Bhattacharjee}.}
  \bibinfo{year}{2018}\natexlab{}.
\newblock \showarticletitle{{Internet Anycast: Performance, Problems, and
  Potential}}. In \bibinfo{booktitle}{\emph{Proc. of the ACM SIGCOMM
  Conference}}. \bibinfo{publisher}{{ACM}}, \bibinfo{address}{Budapest,
  Hungary}, \bibinfo{pages}{59--73}.
\newblock
\urldef\tempurl%
\url{https://doi.org/10.1145/3230543.3230547}
\showDOI{\tempurl}


\bibitem[Liu and Liu(2009)]%
        {liu09}
\bibfield{author}{\bibinfo{person}{Baoding Liu} {and} \bibinfo{person}{Baoding
  Liu}.} \bibinfo{year}{2009}\natexlab{}.
\newblock \showarticletitle{{Facility location problem}}.
\newblock \bibinfo{journal}{\emph{Theory and practice of uncertain
  programming}} (\bibinfo{year}{2009}), \bibinfo{pages}{157--165}.
\newblock


\bibitem[Livadariu et~al\mbox{.}(2020)]%
        {Livadariu20}
\bibfield{author}{\bibinfo{person}{Ioana Livadariu}, \bibinfo{person}{Thomas
  Dreibholz}, \bibinfo{person}{Anas~Saeed Al-Selwi}, \bibinfo{person}{Haakon
  Bryhni}, \bibinfo{person}{Olav Lysne}, \bibinfo{person}{Steinar
  Bj\o{}rnstad}, {and} \bibinfo{person}{Ahmed Elmokashfi}.}
  \bibinfo{year}{2020}\natexlab{}.
\newblock \showarticletitle{{On the Accuracy of Country-Level IP Geolocation}}.
  In \bibinfo{booktitle}{\emph{Proceedings of the Applied Networking Research
  Workshop}} (Virtual Event, Spain) \emph{(\bibinfo{series}{ANRW '20})}.
  \bibinfo{publisher}{Association for Computing Machinery},
  \bibinfo{address}{New York, NY, USA}, \bibinfo{pages}{67–73}.
\newblock
\showISBNx{9781450380393}
\urldef\tempurl%
\url{https://doi.org/10.1145/3404868.3406664}
\showDOI{\tempurl}


\bibitem[Marianov and Serra(2011)]%
        {marianov2011median}
\bibfield{author}{\bibinfo{person}{Vladimir Marianov} {and}
  \bibinfo{person}{Daniel Serra}.} \bibinfo{year}{2011}\natexlab{}.
\newblock \showarticletitle{{Median Problems in Networks: Introduction}}.
\newblock In \bibinfo{booktitle}{\emph{Foundations of Location Analysis}},
  \bibfield{editor}{\bibinfo{person}{H.~A. Eiselt} {and}
  \bibinfo{person}{Vladimir Marianov}} (Eds.). \bibinfo{publisher}{Springer},
  \bibinfo{pages}{31--43}.
\newblock
\showISBNx{978-1-4419-7571-3}


\bibitem[Meta({[n.\,d.]})]%
        {MetaDatacenters}
\bibfield{author}{\bibinfo{person}{Meta}.} \bibinfo{year}{[n.\,d.]}\natexlab{}.
\newblock \bibinfo{title}{{U.S. datacenter fleet}}.
\newblock
  \bibinfo{howpublished}{\url{https://datacenters.atmeta.com/us-locations/}}.
\newblock


\bibitem[Mitchell({[n.\,d.]})]%
        {Traceroutes}
\bibfield{author}{\bibinfo{person}{Scott Mitchell}.}
  \bibinfo{year}{[n.\,d.]}\natexlab{}.
\newblock \bibinfo{title}{{How to Read a Traceroute}}.
\newblock
  \bibinfo{howpublished}{\url{https://www.inmotionhosting.com/support/server/ssh/read-traceroute/}}.
\newblock


\bibitem[Narayanan et~al\mbox{.}(2020a)]%
        {narayanan2020first}
\bibfield{author}{\bibinfo{person}{Arvind Narayanan}, \bibinfo{person}{Eman
  Ramadan}, \bibinfo{person}{Jason Carpenter}, \bibinfo{person}{Qingxu Liu},
  \bibinfo{person}{Yu Liu}, \bibinfo{person}{Feng Qian}, {and}
  \bibinfo{person}{Zhi-Li Zhang}.} \bibinfo{year}{2020}\natexlab{a}.
\newblock \showarticletitle{{A first look at commercial 5G performance on
  smartphones}}. In \bibinfo{booktitle}{\emph{Proceedings of The Web
  Conference}}. \bibinfo{pages}{894--905}.
\newblock


\bibitem[Narayanan et~al\mbox{.}(2020b)]%
        {narayanan2020lumos5g}
\bibfield{author}{\bibinfo{person}{Arvind Narayanan}, \bibinfo{person}{Eman
  Ramadan}, \bibinfo{person}{Rishabh Mehta}, \bibinfo{person}{Xinyue Hu},
  \bibinfo{person}{Qingxu Liu}, \bibinfo{person}{Rostand~AK Fezeu},
  \bibinfo{person}{Udhaya~Kumar Dayalan}, \bibinfo{person}{Saurabh Verma},
  \bibinfo{person}{Peiqi Ji}, \bibinfo{person}{Tao Li}, {et~al\mbox{.}}}
  \bibinfo{year}{2020}\natexlab{b}.
\newblock \showarticletitle{{Lumos5G: Mapping and predicting commercial mmWave
  5G throughput}}. In \bibinfo{booktitle}{\emph{Proceedings of the ACM internet
  measurement conference}}. \bibinfo{pages}{176--193}.
\newblock


\bibitem[Narayanan et~al\mbox{.}(2022)]%
        {narayanan2022comparative}
\bibfield{author}{\bibinfo{person}{Arvind Narayanan},
  \bibinfo{person}{Muhammad~Iqbal Rochman}, \bibinfo{person}{Ahmad Hassan},
  \bibinfo{person}{Bariq~S Firmansyah}, \bibinfo{person}{Vanlin Sathya},
  \bibinfo{person}{Monisha Ghosh}, \bibinfo{person}{Feng Qian}, {and}
  \bibinfo{person}{Zhi-Li Zhang}.} \bibinfo{year}{2022}\natexlab{}.
\newblock \showarticletitle{{A comparative measurement study of commercial 5G
  mmWave deployments}}. In \bibinfo{booktitle}{\emph{IEEE INFOCOM 2022-IEEE
  Conference on Computer Communications}}. IEEE.
\newblock


\bibitem[{Narayanan, Arvind and Zhang, Xumiao and Zhu, Ruiyang and Hassan,
  Ahmad and Jin, Shuowei and Zhu, Xiao and Zhang, Xiaoxuan and Rybkin, Denis
  and Yang, Zhengxuan and Mao, Zhuoqing Morley and others}(2021)]%
        {narayanan2021variegated}
\bibfield{author}{\bibinfo{person}{{Narayanan, Arvind and Zhang, Xumiao and
  Zhu, Ruiyang and Hassan, Ahmad and Jin, Shuowei and Zhu, Xiao and Zhang,
  Xiaoxuan and Rybkin, Denis and Yang, Zhengxuan and Mao, Zhuoqing Morley and
  others}}.} \bibinfo{year}{2021}\natexlab{}.
\newblock \showarticletitle{{A variegated look at 5G in the wild: performance,
  power, and QoE implications}}. In \bibinfo{booktitle}{\emph{Proceedings of
  the 2021 ACM SIGCOMM 2021 Conference}}. \bibinfo{pages}{610--625}.
\newblock


\bibitem[Ni et~al\mbox{.}(2023)]%
        {ni2023polycorn}
\bibfield{author}{\bibinfo{person}{Yunzhe Ni}, \bibinfo{person}{Feng Qian},
  \bibinfo{person}{Taide Liu}, \bibinfo{person}{Yihua Cheng},
  \bibinfo{person}{Zhiyao Ma}, \bibinfo{person}{Jing Wang},
  \bibinfo{person}{Zhongfeng Wang}, \bibinfo{person}{Gang Huang},
  \bibinfo{person}{Xuanzhe Liu}, {and} \bibinfo{person}{Chenren Xu}.}
  \bibinfo{year}{2023}\natexlab{}.
\newblock \showarticletitle{{$\{$POLYCORN$\}$: Data-driven cross-layer
  multipath networking for high-speed railway through composable
  schedulerlets}}. In \bibinfo{booktitle}{\emph{20th USENIX Symposium on
  NSDI}}. \bibinfo{pages}{1325--1340}.
\newblock


\bibitem[Nur(2023)]%
        {Yasin2023}
\bibfield{author}{\bibinfo{person}{Abdullah~Yasin Nur}.}
  \bibinfo{year}{2023}\natexlab{}.
\newblock \showarticletitle{{Accuracy and Coverage Analysis of IP Geolocation
  Databases}}. In \bibinfo{booktitle}{\emph{2023 International Balkan
  Conference on Communications and Networking (BalkanCom)}}.
  \bibinfo{pages}{1--6}.
\newblock
\urldef\tempurl%
\url{https://doi.org/10.1109/BalkanCom58402.2023.10167899}
\showDOI{\tempurl}


\bibitem[OpenCellID({[n.\,d.]})]%
        {OpenCellID}
\bibfield{author}{\bibinfo{person}{OpenCellID}.}
  \bibinfo{year}{[n.\,d.]}\natexlab{}.
\newblock \bibinfo{title}{{OpenCellID database}}.
\newblock \bibinfo{howpublished}{\url{https://www.opencellid.org/}}.
\newblock


\bibitem[Parastar et~al\mbox{.}(2023)]%
        {parastar2023spotlight}
\bibfield{author}{\bibinfo{person}{Paniz Parastar}, \bibinfo{person}{Andra
  Lutu}, \bibinfo{person}{{\"O}zg{\"u} Alay}, \bibinfo{person}{Giuseppe Caso},
  {and} \bibinfo{person}{Diego Perino}.} \bibinfo{year}{2023}\natexlab{}.
\newblock \showarticletitle{{Spotlight on 5G: Performance, device evolution and
  challenges from a mobile operator perspective}}. In
  \bibinfo{booktitle}{\emph{IEEE INFOCOM 2023}}. IEEE, \bibinfo{pages}{1--10}.
\newblock


\bibitem[PeeringDB(2024)]%
        {PeeringDBGoogle24a}
\bibfield{author}{\bibinfo{person}{PeeringDB}.}
  \bibinfo{year}{2024}\natexlab{}.
\newblock \bibinfo{title}{Google Interconnection Facilities}.
\newblock \bibinfo{howpublished}{\url{https://www.peeringdb.com/net/433}}.
\newblock
\newblock
\shownote{Accessed 2024-07-26.}.


\bibitem[Ramanathan and Abdu~Jyothi(2023)]%
        {Alagappan23a}
\bibfield{author}{\bibinfo{person}{Alagappan Ramanathan} {and}
  \bibinfo{person}{Sangeetha Abdu~Jyothi}.} \bibinfo{year}{2023}\natexlab{}.
\newblock \showarticletitle{Nautilus: A Framework for Cross-Layer Cartography
  of Submarine Cables and IP Links}.
\newblock \bibinfo{journal}{\emph{Proc. ACM Meas. Anal. Comput. Syst.}}
  \bibinfo{volume}{7}, \bibinfo{number}{3}, Article \bibinfo{articleno}{46}
  (\bibinfo{date}{Dec.} \bibinfo{year}{2023}), \bibinfo{numpages}{34}~pages.
\newblock
\urldef\tempurl%
\url{https://doi.org/10.1145/3626777}
\showDOI{\tempurl}


\bibitem[Rosen et~al\mbox{.}(2014)]%
        {rosen2014discovering}
\bibfield{author}{\bibinfo{person}{Sanae Rosen}, \bibinfo{person}{Haokun Luo},
  \bibinfo{person}{Qi~Alfred Chen}, \bibinfo{person}{Z~Morley Mao},
  \bibinfo{person}{Jie Hui}, \bibinfo{person}{Aaron Drake}, {and}
  \bibinfo{person}{Kevin Lau}.} \bibinfo{year}{2014}\natexlab{}.
\newblock \showarticletitle{{Discovering fine-grained RRC state dynamics and
  performance impacts in cellular networks}}. In
  \bibinfo{booktitle}{\emph{Proceedings of the 20th annual international
  conference on Mobile computing and networking}}. \bibinfo{pages}{177--188}.
\newblock


\bibitem[Shkilko and Sokolov(2020)]%
        {Shkilko20a}
\bibfield{author}{\bibinfo{person}{Andriy Shkilko} {and}
  \bibinfo{person}{Konstantin Sokolov}.} \bibinfo{year}{2020}\natexlab{}.
\newblock \showarticletitle{{Every Cloud Has a Silver Lining: Fast Trading,
  Microwave Connectivity, and Trading Costs}}.
\newblock \bibinfo{journal}{\emph{Journal of Finance}} \bibinfo{volume}{75},
  \bibinfo{number}{6} (\bibinfo{date}{Dec.} \bibinfo{year}{2020}),
  \bibinfo{pages}{2899--2927}.
\newblock
\urldef\tempurl%
\url{https://doi.org/10.1111/jofi.12969}
\showDOI{\tempurl}


\bibitem[Singla et~al\mbox{.}(2014)]%
        {Singla14b}
\bibfield{author}{\bibinfo{person}{Ankit Singla}, \bibinfo{person}{Balakrishnan
  Chandrasekaran}, \bibinfo{person}{P.~Brighten Godfrey}, {and}
  \bibinfo{person}{Bruce Maggs}.} \bibinfo{year}{2014}\natexlab{}.
\newblock \showarticletitle{{The {Internet} at the Speed of Light}}. In
  \bibinfo{booktitle}{\emph{Proceedings of the ACM Workshop on Hot Topics in
  Networks}}. \bibinfo{publisher}{{ACM}}, \bibinfo{address}{Los Angeles, CA,
  USA}.
\newblock
\urldef\tempurl%
\url{https://doi.org/10.1145/2670518.2673876}
\showDOI{\tempurl}


\bibitem[Spring et~al\mbox{.}(2023)]%
        {Spring23}
\bibfield{author}{\bibinfo{person}{Neil Spring}, \bibinfo{person}{Ratul
  Mahajan}, {and} \bibinfo{person}{Thomas Anderson}.}
  \bibinfo{year}{2003}\natexlab{}.
\newblock \showarticletitle{{The causes of path inflation}}. In
  \bibinfo{booktitle}{\emph{Proceedings of the 2003 Conference on Applications,
  Technologies, Architectures, and Protocols for Computer Communications}}
  (Karlsruhe, Germany) \emph{(\bibinfo{series}{SIGCOMM '03})}.
  \bibinfo{publisher}{Association for Computing Machinery},
  \bibinfo{address}{New York, NY, USA}, \bibinfo{pages}{113–124}.
\newblock
\showISBNx{1581137354}
\urldef\tempurl%
\url{https://doi.org/10.1145/863955.863970}
\showDOI{\tempurl}


\bibitem[Sundberg et~al\mbox{.}(2024)]%
        {sundberg24}
\bibfield{author}{\bibinfo{person}{Simon Sundberg}, \bibinfo{person}{Anna
  Brunstrom}, \bibinfo{person}{Simone Ferlin-Reiter}, \bibinfo{person}{Toke
  H{\o}iland-J{\o}rgensen}, {and} \bibinfo{person}{Robert Chac{\'o}n}.}
  \bibinfo{year}{2024}\natexlab{}.
\newblock \showarticletitle{{Measuring Network Latency from a Wireless ISP:
  Variations Within and Across Subnets}}. In
  \bibinfo{booktitle}{\emph{Proceedings of the 2024 ACM on Internet Measurement
  Conference}}. \bibinfo{pages}{29--43}.
\newblock


\bibitem[Varvello and Zaki(2023)]%
        {varvello23}
\bibfield{author}{\bibinfo{person}{Matteo Varvello} {and}
  \bibinfo{person}{Yasir Zaki}.} \bibinfo{year}{2023}\natexlab{}.
\newblock \showarticletitle{{A worldwide look into mobile access networks
  through the eyes of amigos}}. In \bibinfo{booktitle}{\emph{2023 7th Network
  Traffic Measurement and Analysis Conference (TMA)}}. IEEE,
  \bibinfo{pages}{1--10}.
\newblock


\bibitem[Wang et~al\mbox{.}({[n.\,d.]})]%
        {wangcalypso}
\bibfield{author}{\bibinfo{person}{Caleb~J Wang}, \bibinfo{person}{Ying Zhang},
  \bibinfo{person}{Qianli Dong}, \bibinfo{person}{Esteban Carisimo},
  \bibinfo{person}{Ramakrishnan Durairajan}, {and}
  \bibinfo{person}{Fabi{\'a}n~E Bustamante}.}
  \bibinfo{year}{[n.\,d.]}\natexlab{}.
\newblock \showarticletitle{Calypso’s Voyage: Charting Traceroute Paths to
  Submarine Cables}.
\newblock  (\bibinfo{year}{[n.\,d.]}).
\newblock


\bibitem[Wang et~al\mbox{.}(2019)]%
        {wang2019active}
\bibfield{author}{\bibinfo{person}{Jing Wang}, \bibinfo{person}{Yufan Zheng},
  \bibinfo{person}{Yunzhe Ni}, \bibinfo{person}{Chenren Xu},
  \bibinfo{person}{Feng Qian}, \bibinfo{person}{Wangyang Li},
  \bibinfo{person}{Wantong Jiang}, \bibinfo{person}{Yihua Cheng},
  \bibinfo{person}{Zhuo Cheng}, \bibinfo{person}{Yuanjie Li}, {et~al\mbox{.}}}
  \bibinfo{year}{2019}\natexlab{}.
\newblock \showarticletitle{{An active-passive measurement study of TCP
  performance over LTE on high-speed rails}}. In \bibinfo{booktitle}{\emph{The
  25th Annual International Conference on Mobile Computing and Networking}}.
\newblock


\bibitem[Xu et~al\mbox{.}(2020)]%
        {xu2020understanding}
\bibfield{author}{\bibinfo{person}{Dongzhu Xu}, \bibinfo{person}{Anfu Zhou},
  \bibinfo{person}{Xinyu Zhang}, \bibinfo{person}{Guixian Wang},
  \bibinfo{person}{Xi Liu}, \bibinfo{person}{Congkai An},
  \bibinfo{person}{Yiming Shi}, \bibinfo{person}{Liang Liu}, {and}
  \bibinfo{person}{Huadong Ma}.} \bibinfo{year}{2020}\natexlab{}.
\newblock \showarticletitle{{Understanding operational 5G: A first measurement
  study on its coverage, performance and energy consumption}}. In
  \bibinfo{booktitle}{\emph{Proc. of the ACM Special Interest Group on Data
  Communication on the applications, technologies, architectures, and protocols
  for computer communication}}. \bibinfo{pages}{479--494}.
\newblock


\bibitem[Yang et~al\mbox{.}(2022)]%
        {yang2022mobile}
\bibfield{author}{\bibinfo{person}{Xinlei Yang}, \bibinfo{person}{Hao Lin},
  \bibinfo{person}{Zhenhua Li}, \bibinfo{person}{Feng Qian},
  \bibinfo{person}{Xingyao Li}, \bibinfo{person}{Zhiming He},
  \bibinfo{person}{Xudong Wu}, \bibinfo{person}{Xianlong Wang},
  \bibinfo{person}{Yunhao Liu}, \bibinfo{person}{Zhi Liao}, {et~al\mbox{.}}}
  \bibinfo{year}{2022}\natexlab{}.
\newblock \showarticletitle{{Mobile Access Bandwidth in Practice:
  Measurement}}.
\newblock \bibinfo{journal}{\emph{Analysis, and Implications. In SIGCOMM}}
  \bibinfo{volume}{2}, \bibinfo{number}{2} (\bibinfo{year}{2022}),
  \bibinfo{pages}{3--1}.
\newblock


\bibitem[Zarifis et~al\mbox{.}(2014)]%
        {zarifis2014diagnosing}
\bibfield{author}{\bibinfo{person}{Kyriakos Zarifis}, \bibinfo{person}{Tobias
  Flach}, \bibinfo{person}{Srikanth Nori}, \bibinfo{person}{David Choffnes},
  \bibinfo{person}{Ramesh Govindan}, \bibinfo{person}{Ethan Katz-Bassett},
  \bibinfo{person}{Z~Morley Mao}, {and} \bibinfo{person}{Matt Welsh}.}
  \bibinfo{year}{2014}\natexlab{}.
\newblock \showarticletitle{Diagnosing path inflation of mobile client
  traffic}. In \bibinfo{booktitle}{\emph{International Conference on Passive
  and Active Network Measurement}}. Springer, \bibinfo{pages}{23--33}.
\newblock


\bibitem[Zhang et~al\mbox{.}(2021)]%
        {Zhang21b}
\bibfield{author}{\bibinfo{person}{Zesen Zhang}, \bibinfo{person}{Alexander
  Marder}, \bibinfo{person}{Ricky Mok}, \bibinfo{person}{Bradley Huffaker},
  \bibinfo{person}{Matthew Luckie}, \bibinfo{person}{K~C Claffy}, {and}
  \bibinfo{person}{Aaron Schulman}.} \bibinfo{year}{2021}\natexlab{}.
\newblock \showarticletitle{{Inferring Regional Access Network Topologies:
  Methods and Applications}}. In \bibinfo{booktitle}{\emph{Proceedings of the
  ACM Internet Measurement Conference}}. \bibinfo{publisher}{{ACM}},
  \bibinfo{address}{Virtual Event}.
\newblock
\urldef\tempurl%
\url{https://doi.org/10.1145/3487552.3487812}
\showDOI{\tempurl}


\bibitem[Zhang et~al\mbox{.}(2019)]%
        {Zhang19b}
\bibfield{author}{\bibinfo{person}{Zhehui Zhang}, \bibinfo{person}{Shu Shi},
  \bibinfo{person}{Varun Gupta}, {and} \bibinfo{person}{Rittwik Jana}.}
  \bibinfo{year}{2019}\natexlab{}.
\newblock \showarticletitle{{Analysis of Cellular Network Latency for
  Edge-Based Remote Rendering Streaming Applications}}. In
  \bibinfo{booktitle}{\emph{Proceedings of the ACM SIGCOMM 2019 Workshop on
  Networking for Emerging Applications and Technologies}} (Beijing, China)
  \emph{(\bibinfo{series}{NEAT'19})}. \bibinfo{publisher}{Association for
  Computing Machinery}, \bibinfo{address}{New York, NY, USA},
  \bibinfo{pages}{8–14}.
\newblock
\showISBNx{9781450368766}
\urldef\tempurl%
\url{https://doi.org/10.1145/3341558.3342199}
\showDOI{\tempurl}


\bibitem[Zmys{\l}owski and Kelner(2024)]%
        {Zmyslowski24}
\bibfield{author}{\bibinfo{person}{Dariusz Zmys{\l}owski} {and}
  \bibinfo{person}{Jan~M Kelner}.} \bibinfo{year}{2024}\natexlab{}.
\newblock \showarticletitle{{Mobile Network Operators’ Assessment Based on
  Drive-Test Campaign in Urban Area for iPerf Scenario}}.
\newblock \bibinfo{journal}{\emph{Applied Sciences}} \bibinfo{volume}{14},
  \bibinfo{number}{3} (\bibinfo{year}{2024}), \bibinfo{pages}{1268}.
\newblock

\bibitem[Ghoshal et~al\mbox{.}(2025)]%
        {Ghoshal25ReplicationData}
\bibfield{author}{\bibinfo{person}{Moinak Ghoshal},
  \bibinfo{person}{Omar Basit},
  \bibinfo{person}{Imran Khan},
  \bibinfo{person}{Z.~Jonny Kong},
  \bibinfo{person}{Sizhe Wang},
  \bibinfo{person}{Yufei Feng},
  \bibinfo{person}{Phuc Dinh},
  \bibinfo{person}{Y.~Charlie Hu}, {and}
  \bibinfo{person}{Dimitrios Koutsonikolas}.}
  \bibinfo{year}{2025}\natexlab{}.
\newblock \bibinfo{title}{Data and scripts from the paper
  ``Replication: Performance of Cellular Networks on the Wheels''}.
\newblock \bibinfo{howpublished}{GitHub repository}.
\newblock
\shownote{Measurements collected November 1--8, 2024}.
\newblock
\urldef\tempurl%
\url{https://github.com/NUWiNS/imc2025-replication-cellular-network-performance-on-wheels-data}
\showURL{%
\tempurl}

\bibitem[Kulkarni and Heidemann(2026)]%
        {Kulkarni26c}
\bibfield{author}{\bibinfo{person}{Shefali Tushar Kulkarni} {and}
  \bibinfo{person}{John Heidemann}.}
  \bibinfo{year}{2026}\natexlab{}.
\newblock \bibinfo{title}{Data from the paper ``Bounding Mobile
  Latency from Network Interconnections''}.
\newblock \bibinfo{howpublished}{Web page}.
  \bibinfo{date}{October} \bibinfo{year}{2026}.
\newblock
\urldef\tempurl%
\url{https://ant.isi.edu/datasets/mobile_latency}
\showURL{%
\tempurl}


\bibitem[Kulkarni and Heidemann(2026)]%
        {Kulkarni26a}
\bibfield{author}{\bibinfo{person}{Shefali Tushar Kulkarni} {and}
  \bibinfo{person}{John Heidemann}.}
  \bibinfo{year}{2026}\natexlab{}.
\newblock \showarticletitle{Bounding Mobile Latency from Network
  Interconnections}. In
  \bibinfo{booktitle}{\emph{Proceedings of the 21st Asian Internet
  Engineering Conference}} (Taipei, Taiwan)
  \emph{(\bibinfo{series}{AINTEC '26})}.
  \bibinfo{publisher}{Association for Computing Machinery}.
\newblock
\shownote{To appear}.
\newblock
\urldef\tempurl%
\url{https://ant.isi.edu/~johnh/PAPERS/Kulkarni26a.html}
\showURL{%
\tempurl}



\end{thebibliography}

\appendix
\section{Assessing the Role of Distance in Latency Overhead}
\label{app:distance_latency}

To examine whether geographic distance influences exit-path performance,
we analyze the relationship between exit-path distance and the additional exit path latency. 

\begin{figure}
    \centering
    \includegraphics[width=0.9\columnwidth]{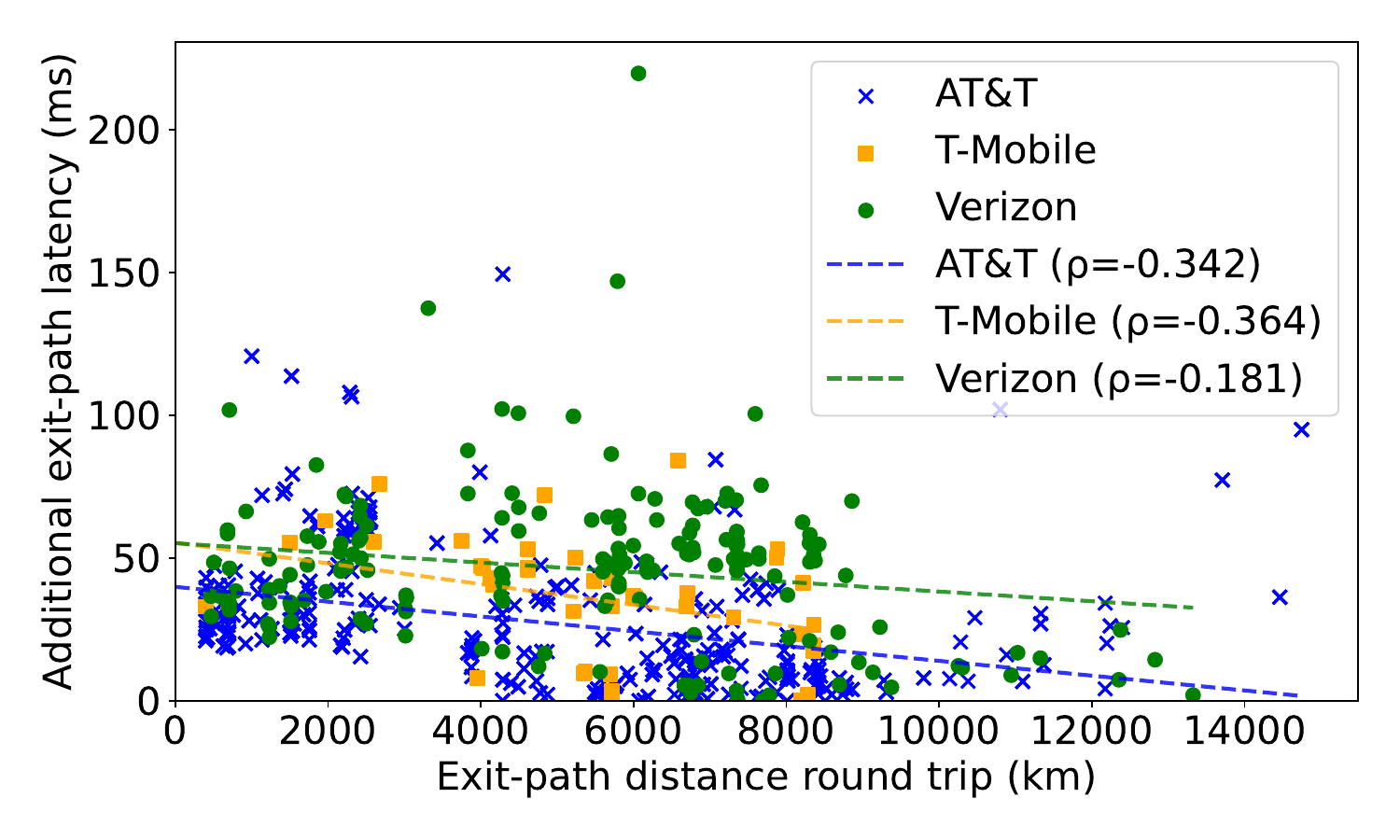}
    \caption{Scatter plot of additional exit-path latency vs. exit-path distance. Each point represents a measurement, with colors and marker shapes indicating the MNO.
}
    \label{fig:additional_exit_latency_vs_distance}
\end{figure}

Additional exit path latency is
defined in \autoref{sec:latency_model}.
\autoref{fig:additional_exit_latency_vs_distance} plots additional exit-path latency against the exit-path distance from the mobile vantage point to the server. 
Because the relationship is not visually apparent from the scatter
plots, 
we quantify the linear association using Pearson's correlation
coefficient ($r$).

All three MNOs exhibit a weak-to-moderate negative correlation between exit-path distance and additional latency: AT\&T has $r=-0.342$,
T-Mobile has
$r=-0.364$, 
and Verizon has $r=-0.181$. 
The dotted lines in
\autoref{fig:additional_exit_latency_vs_distance} show the corresponding linear
trends. 
Contrary to our initial expectation that additional latency would
increase with distance, 
the measurements indicate that shorter exit paths tend
to incur greater additional latency.

One possible explanation is that non-propagation delays, including routing inefficiencies, 
processing, 
and queuing, 
are more prominent for geographically
short paths, 
such as Los Angeles to San Diego. 
However, 
fixed delay alone would
produce an approximately distance-independent absolute overhead rather than a
negative relationship. 
The observed trend may therefore also indicate that
longer exit paths in our measurements tend to encounter more efficient routing
or lower non-propagation delay. 
These correlations describe an association and
do not, 
by themselves, 
establish its cause.

To determine whether this relationship is specific to exit-path distance, 
we repeat the analysis using direct-path distance. 
We find almost no correlation
for AT\&T and Verizon ($r=0.026$ and $r=-0.024$, respectively), 
but a moderate negative correlation for T-Mobile ($r=-0.405$). 
One possible explanation for
T-Mobile is that much of its traffic traverses a distant backbone node in Denver, even when the source is geographically close to San Diego. 
Consequently,
geographic direct-path distance may not accurately represent the distance actually traversed within T-Mobile's network. 

Together, 
these results reinforce
the importance of accounting for MNO exit-point placement and internal routing
when interpreting mobile-network latency.

\section{Better Placement Beats More Exits}
    \label{app:optimizing_n+i_location}

We state that adding exit points provides only modest latency
improvements, 
whereas optimizing the placement of existing exit points provides
substantially greater benefits in \autoref{sec:exit_optimality}.
Here, we present the detailed analysis
to support this finding.

We estimate the potential latency gains as the number
of optimally placed exit points increases. 
Because the optimization ignores
practical constraints such as datacenter availability and rights-of-way, 
these results represent a best-case estimate of the potential improvement rather than an immediately deployable configuration.

\begin{figure}
   \centering
   \includegraphics[width=0.9\linewidth]{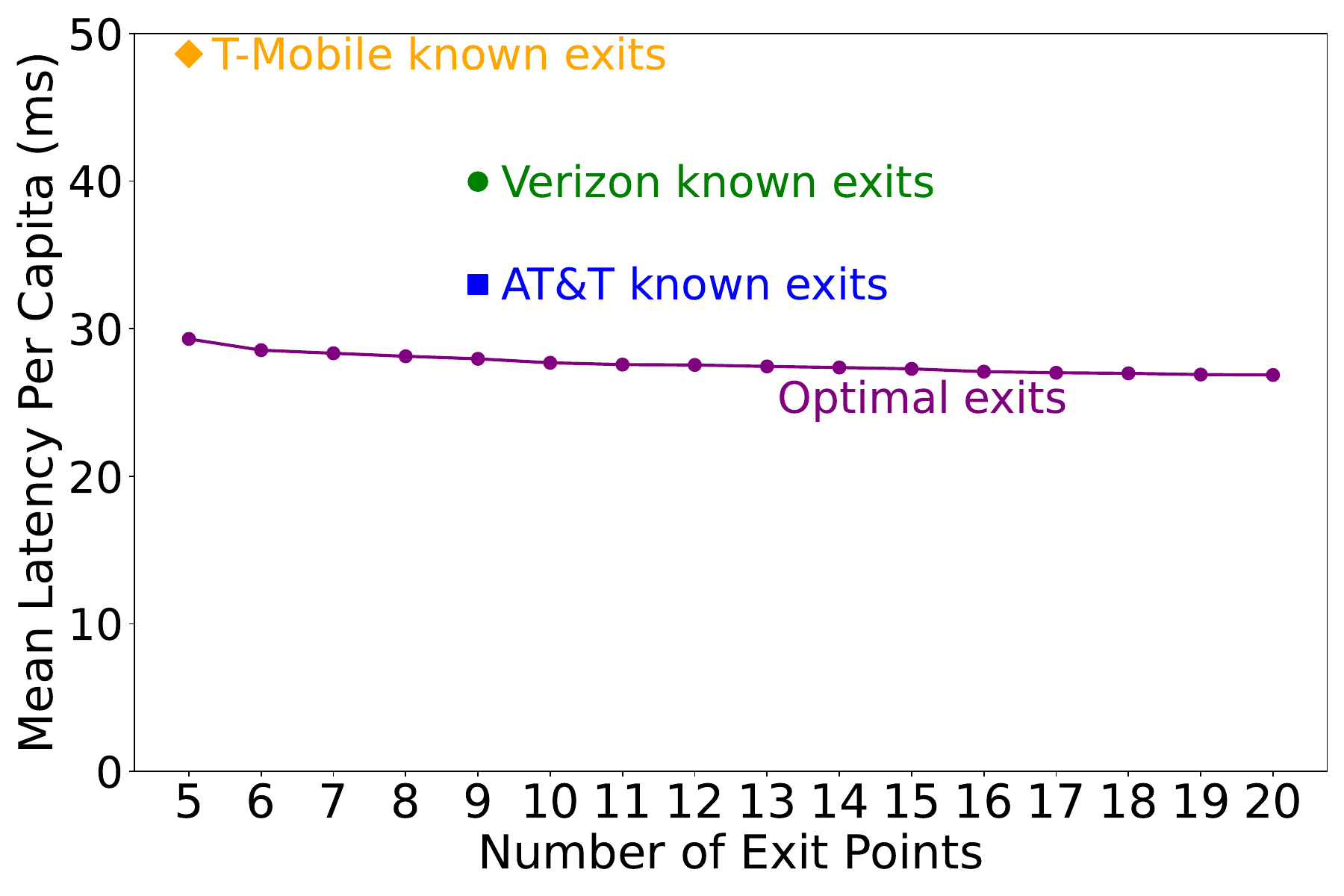}
   \caption{ Optimal vs Known Mean latency per capita and improvement from adding a new exit point in optimal location compared to the previous number of exit points }
   \label{fig:n+i_latency}
\end{figure}

We expect additional exit points to lower mean latency per capita.
\autoref{fig:n+i_latency} shows this metric (purple line) as the number of optimally placed exits increases from 5 (T-Mobile's known count) to 20.
Each new exit yields only modest improvement,
  reducing latency by 1\% or less.
In contrast,
relocating existing exits to optimal positions provides a significant latency reduction.
This is evident by comparing the colored dots (actual MNO exit-point latency) to the purple line (optimal placements).
While our optimal locations ignore practical constraints like datacenter availability or rights-of-way,
they highlight opportunities for feasible adjustments.
In summary, better placement of existing exits offers more benefit than simply adding more.


\end{document}